\RequirePackage{pdf14}
\documentclass[letterpaper]{article}
\pdfoutput=1

\usepackage[usenames,dvipsnames,svgnames,table]{xcolor}
\usepackage{array,setspace,mathrsfs,amsfonts,amsmath,yfonts,dsfont,mathabx,bbm,colonequals,youngtab}
\usepackage{./aux/tikz-cd}
\usepackage{./aux/jheppub}

\newtheorem{con}{Conjecture}

\def \beg#1{\begin{#1}} 
\def \be{\beg{equation}}
\def \bea{\beg{eqnarray}}
\def \eea{\end{eqnarray}}
\def \ee{\end{equation}}

\newcommand\qq{\mathbbmtt{Q}\,}

\def \PE{\mathrm{P.E.}}
\def \NO#1{\mathop{:}\nolimits\!#1\!\mathop{:}\nolimits}

\def \Qt{\wt\QQ}

\def \Lh{\widehat{L}}

\def \gf{\mf{g}}
\def \uf{\mf{u}}
\def \slf{\mf{sl}}
\def \suf{\mf{su}}
\def \sof{\mf{so}}

\def \uspf{\mf{usp}}
\def \ospf{\mf{osp}}

\def \hhf{\mf{h}}
\def \ceq{\colonequals}

\def \restr#1#2{{\left.\kern-\nulldelimiterspace#1\vphantom{\big|}\right|_{#2}}}

\def \Poincare{Poincar\'e}

\def \ba{{\bf a}}
\def \bb{{\bf b}}

\def \bd{{\bf d}}

\def \bA{{\bf A}}
\def \bB{{\bf B}}
\def \bC{{\bf C}}
\def \bD{{\bf D}}
\def \bOn{{\bf 1}}
\def \bTw{{\bf 2}}
\def \bTh{{\bf 3}}
\def \bFo{{\bf 4}}
\def \bFi{{\bf 5}}
\def \cd{{\tilde c}}

\def \hf{\frac12}

\def \mf{\mathfrak}

\def \nn{\nonumber}
\def \ph{\phantom}
\def \wt{\widetilde}
\def \wh{\widehat}
\def \ol{\overline}
\def \eg{{\it e.g.}}
\def \ie{{\it i.e.}}
\def \cf{{\it cf.}}
\def \Tr{{\rm Tr}}

\def \zb{{\bar z}}

\def \dag{\dagger}

\def \Cb{\mathbb{C}}

\def \Rb{\mathbb{R}}

\def \AA{\mathcal{A}}
\def \BB{\mathcal{B}}
\def \CC{\mathcal{C}}
\def \DD{\mathcal{D}}

\def \HH{\mathcal{H}}
\def \II{\mathcal{I}}

\def \KK{\mathcal{K}}
\def \LL{\mathcal{L}}
\def \MM{\mathcal{M}}
\def \NN{\mathcal{N}}
\def \OO{\mathcal{O}}
\def \PP{\mathcal{P}}
\def \QQ{\mathcal{Q}}
\def \RR{\mathcal{R}}
\def \SS{\mathcal{S}}

\def \WW{\mathcal{W}}
\def \XX{\mathcal{X}}

\def \ZZ{\mathcal{Z}}

\title{
\vspace{1.25in}
\texorpdfstring{$\WW$}{W} symmetry in six dimensions
}

\author[\!1]{Christopher Beem,}
\author[\!1,2,3]{Leonardo Rastelli,}
\author[\!4]{Balt C. van Rees\ph{,}}

\affiliation[1]{Institute for Advanced Study, Einstein Dr., Princeton, NJ 08540, USA}
\affiliation[2]{Kavli Institute for Theoretical Physics, UCSB, Santa Barbara, CA 93106, USA}
\affiliation[3]{C.~N.~Yang Institute for Theoretical Physics, Stony Brook University, Stony Brook, NY 11794-3840, USA}
\affiliation[4]{Theory Group, Physics Department, CERN, CH-1211 Geneva 23, Switzerland}

\preprint{CERN-PH-TH/2014-56}
\bigskip
\abstract{
Six-dimensional conformal field theories with $(2,0)$ supersymmetry are shown to possess a protected sector of operators and observables that are isomorphic to a two-dimensional chiral algebra. We argue that the chiral algebra associated to a $(2,0)$ theory labelled by the simply-laced Lie algebra $\gf$ is precisely the $\WW$ algebra of type $\gf$, for a specific value of the central charge. Simple examples of observables that are made accessible by this correspondence are the three-point functions of half-BPS operators. For the $A_n$ series, we compare our results at large $n$ to those obtained using the holographic dual description and find perfect agreement. We further find protected chiral algebras that appear on the worldvolumes of codimension two defects in $(2,0)$ SCFTs. This construction has likely implications for understanding the microscopic origin of the AGT correspondence.
}


\notoc
\begin{document}
\maketitle
\vfill\eject
%

\section{Introduction and summary}

It has recently been observed that any four-dimensional conformal field theory with extended supersymmetry has a protected sector that is isomorphic to a two-dimensional chiral algebra \cite{Beem:2013sza}. The existence of such a sector leads to a wide variety of insights, including new unitarity bounds and powerful organizing principles that underlie the spectrum of BPS operators of such theories. Furthermore, the constraints of crossing symmetry are eminently tractable for correlation functions in this subsector, and solving the ``mini-bootstrap'' problem associated with these constraints is an important preliminary step towards implementing the full numerical bootstrap program for unprotected correlation functions in such theories. An obvious question that presents itself is whether such a structure can be reproduced in superconformal field theories (SCFTs) in spacetimes of dimension $d\neq4$.

The arguments presented in the four-dimensional case were fairly general, with the existence of a protected chiral algebra following entirely from the existence of an $\suf(1,1|2)$ superconformal subalgebra of the full superconformal algebra for which the $\suf(1,1)$ subalgebra acts as anti-holomorphic M\"obius transformations on some fixed plane. A similar subsector will consequently exist in any theory for which the superconformal algebra includes such a subalgebra. A quick survey of the available superconformal algebras \cite{Kac:1977qb,Nahm:1977tg} leads to a rather short list of possibilities:
\begin{enumerate}
\item[I] 	\qquad	$\suf(2,2|2)$		: $\NN=2$ in $d=4$.
\item[II]   \qquad	$\suf(2,2|4)$		: $\NN=4$ in $d=4$.
\item[III] 	\qquad	$\ospf(8^\star|4)$	: $\NN=(2,0)$ in $d=6$.
\item[IV] 	\qquad	$\suf(1,1|2)$		: ``Small'' $\NN=(0,4)$ and $\NN=(4,4)$ in $d=2$.
\end{enumerate}
The first two entries on this list were the subject of \cite{Beem:2013sza}. In this work we explore the third.

The six-dimensional case holds particular interest since six-dimensional $(2,0)$ SCFTs remain quite mysterious. To the best of our knowledge, no correlation functions have been computed in these theories except in the free case, or indirectly for the $A_n$ theories at large $n$ by means of the AdS/CFT correspondence \cite{Corrado:1999pi, Bastianelli:1999en}. This makes the presence of a solvable subsector all the more interesting, as the structure of the computable correlators may hold some clues about the right language with which to describe $(2,0)$ SCFTs more generally.

The appearance of chiral algebras in the context of the six-dimensional $(2,0)$ theories does not come as a complete surprise. The AGT correspondence \cite{Alday:2009aq, Wyllard:2009hg} relates instanton partition functions of four-dimensional theories of class $\SS$ \cite{Gaiotto:2009we} to Toda correlators, suggesting a deep connection between $(2,0)$ SCFTs and chiral algebras. More precisely, there should be a connection between the $(2,0)$ theory labelled by the simply laced Lie algebra $\gf$ and the chiral algebra $\WW_{\gf}$. However, the microscopic origin of this symmetry has so far remained unclear. Our main result is that the protected chiral algebra associated to the $(2,0)$ SCFT of type $\gf$ is precisely the $\WW_{\gf}$ algebra! In this context, the generating currents of $\WW_{\gf}$ arise very concretely as cohomology classes of half-BPS local operators in the SCFT. This observation seems a likely starting point for a truly microscopic understanding of the AGT correspondence.

Our analysis involves a few technicalities, but the essential argument is not difficult to summarize. As in \cite{Beem:2013sza}, we identify a privileged set of BPS operators that is closed under the operator product expansion. These operators are defined by passing to the cohomology of a certain nilpotent supercharge $\qq$. The requirement that a local operator be annihilated by this supercharge restricts it to lie on a fixed plane $\mathbb{R}^2 \subset \mathbb{R}^6$. The space-time dependence of a $\qq$-closed operator within the fixed plane is also slightly unusual: its orientation in R-symmetry space is correlated with its position on the plane. Concretely, if $(z,\zb)$ are complex coordinate on the plane, the schematic form of a $\qq$-closed operator is
\begin{equation}\label{eq:schematic_twisted_translation}
\OO(z,\zb) \colonequals u_\II(\zb) \OO^\II(z,\zb)\,,
\end{equation}
where $\OO^\II(z,\zb)$ is a conventional local operator that obeys a suitable BPS condition. The index $\II$ runs over the components of a finite-dimensional irreducible representation of the $\sof(5)$ R-symmetry, and  $u_\II(\zb)$ are simple functions of $\zb$. (The precise form is dictated by ``twisting'' the right-moving $\overline{\slf(2)}$ M\"obius symmetry acting on $\zb$ by an $\sof(3)_R$ subgroup of $\sof(5)_R$.) The crucial point of this construction is that the anti-holomorphic position-dependence of such an operator is $\qq$-exact, meaning that its cohomology class depends on the insertion point meromorphically,
\begin{equation}\label{eq:intro_chiral_map}
[\OO(z,\zb)]_\qq~~~\leadsto~~~\OO(z)~.
\end{equation}
Consequently, correlation functions of these twisted operators are meromorphic functions of the insertion points, and as such they inherit the structure of a two-dimensional chiral algebra.

This formal construction associates a chiral algebra to any $(2,0)$ SCFT. Since chiral algebras are very rigid structures, we may hope to completely characterize the ones associated to the known $(2,0)$ theories by leveraging a minimal amount of physical data as input. In particular, the spectrum of half-BPS operators provides a useful starting point for the analysis. Recall that half-BPS operators of a $(2,0)$ theory sit in rank $k$ traceless symmetric tensor representations of $\sof(5)_R$, and have conformal dimension $\Delta = 2 k$. The highest-weight states of these $\sof(5)_R$ representations form a ring. Our starting postulate (well-motivated from several viewpoints \cite{Aharony:1997th,Bhattacharyya:2007sa}) is that this ring is freely generated by a set of elements in one-to-one correspondence with the Casimir invariants of $\gf$ -- in other words the ranks $\{k_i\}$ of the generators of the half-BPS ring coincide with the orders of the Casimir invariants of $\gf$. The cohomological construction maps each of these generators to a generator of the chiral algebra with spin $k_i$. For example, each $(2,0)$ theory contains a single half-BPS operator with $k=2$. This is the superconformal primary of the stress-tensor multiplet. This operator is mapped to a spin-two chiral operator, which plays the role of a holomorphic stress-tensor in the chiral algebra. Higher-rank generators of the half-BPS ring map to higher-spin currents of the chiral algebra. Another piece of information that is not hard to recover is the central charge of the Virasoro symmetry associated with the two-dimensional stress tensor. This can be read off from the appropriately normalized two-point function of half-BPS operators, which in turn is proportional to certain coefficients in the six-dimensional Weyl anomaly.

A natural conjecture is that the generators arising from the half-BPS ring are the \emph{complete} set of generators of the chiral algebra. This guess passes the following non-trivial test. On general grounds, one can argue that the character of the chiral algebra is equal to a certain limit of the superconformal index of the parent $(2,0)$ theory. Precisely this limit has been studied in \cite{Kim:2012ava,Kim:2013nva}, where a simple expression was proposed for the case of the $A_n$ theory. That expression takes precisely the form one would expect for a chiral algebra for which the half-BPS generators are the only generators.

All in all, we are led to the following conjecture:

\begin{con}[Bulk chiral algebra]
The protected chiral algebra of the six-dimensional $(2,0)$ superconformal theory of type $\gf=\{A_n,D_n,E_n\}$ is isomorphic to the $\WW_{\gf}$ chiral algebra with central charge
	\begin{equation}\label{eq:intro_bulk_central_charge}
	c_{2d}=4d_{\gf}h^{\vee}_{\gf}+r_{\gf}~.\nonumber
	\end{equation}
\end{con}

Even after making the above assumptions, our argument falls short of a general proof of this conjecture, as we are not aware of a uniqueness theorem for $\WW$ algebras with the same set of generators as $\WW_\gf$. \emph{A priori}, the associativity constraints may admit multiple solutions for the singular OPE coefficients, leading to inequivalent chiral algebras with identical sets of generators. In simple cases such as $\gf = A_1, A_2$, which correspond to the Virasoro and to the Zamolodchikov $\WW_3$ algebras respectively, it is easy to prove that the crossing symmetry relations admit a unique solution. This is also known to be the case for the $A_3$ and $A_5$ theories \cite{Blumenhagen:1990jv,Hornfeck:1992tm}. For the $A_4$ case, it is known that the solution is not unique \cite{Hornfeck:1992tm}, but the additional solution corresponds to a $\WW$ algebra with null states present for generic values of the central charge, which would conflict with the match of the vacuum character with the superconformal index.

This conjecture has immediate implications for the spectrum and interactions of the $(2,0)$ theories. For example, it predicts that the OPE of two half-BPS operators must contain an infinite tower of protected multiplets obeying certain semi-shortening conditions, with calculable OPE coefficients. This is essential information in setting up the conformal  bootstrap program \cite{Rattazzi:2008pe} for $(2,0)$ theories, along the lines taken in \cite{Beem:2013qxa} for $\NN=4$ SCFTs in four dimensions. The application of bootstrap methods to $(2,0)$ SCFTs will be the subject of a forthcoming publication \cite{forthcoming_20}.

As an illustration of the kind of information that can be extracted from the chiral algebra, we consider three-point functions of half-BPS operators. The $\WW_\gf$ algebra computes them exactly, for any $\gf$. Specializing to $\gf = A_n$, we take the large $n$ limit (for fixed operator dimensions) of the three-point couplings calculated from the chiral algebra, and compare them with the holographic prediction computed using eleven-dimensional supergravity.\footnote{Note that in contrast to the analogous comparison for $\NN=4$ supersymmetric Yang-Mills theory in four dimensions \cite{Freedman:1998tz,Lee:1998bxa} -- wherein a non-renormalization theorem \cite{Baggio:2012rr} allows the correlators in question to be computed at weak 't Hooft coupling and compared to the supergravity computation at strong coupling -- there has heretofore been no independent calculation of these three point functions even at large $n$.} We find an exact match. The agreement of these two completely different, technically very involved calculations is quite miraculous and constitutes strong evidence for our conjecture. More importantly, we now have a procedure for computing half-BPS three-point functions exactly at finite $n$.

\bigskip

With a view towards a microscopic derivation of the AGT correspondence, we also consider superconformal defect theories that preserve an $\suf(2,2|2)$ superalgebra. For any $\gf$, there exists a family of such defects whose members are labelled by embeddings $\rho:\slf(2)\to \gf$~\cite{Gaiotto:2009we}.\footnote{The additional structure associated with outer automorphism twists around the defect (\cf\ \cite{Chacaltana:2012zy}) is left for future work.} In the construction of class $\SS$ theories, these defects create ``punctures'' on the UV curve, with each such defect carrying a global symmetry group equal to the centralizer of $\rho(\slf(2))\subset\gf$. 

The algebraic analysis underlying the existence of a protected chiral algebra in the theory living on these defects is identical to that of \cite{Beem:2013sza}. In particular, the global symmetry of these defects implies the existence of affine currents for the same symmetry algebra in the two-dimensional context. However in contrast to the purely four-dimensional setting, we \emph{do not} expect the chiral algebra associated to defects to include a meromorphic stress tensor, since such an operator is associated with a four-dimensional stress tensor, which will be absent from the defect theories. Though this is a somewhat strange characteristic from a physicist's perspective, such algebras play a major role in the connection between two-dimensional conformal field theory and the geometric Langlands correspondence (see, \eg, \cite{Frenkel:2004jn,Frenkel:2005pa}). We are led to the following natural conjecture, which brings the potential connection to the geometric Langlands into sharp relief:

\begin{con}[Defect chiral algebra]
In the $(2,0)$ theory of type $\gf$, the protected chiral algebra of a codimension two defect labelled by the embedding $\rho$ is isomorphic to the quantum Drinfeld-Sokolov reduction of type $\rho$ of the $\gf$ affine Lie algebra at the critical level,
	\begin{equation}\label{eq:intro_defect_affine_level}
	k_{2d}=-h^\vee~.
	\end{equation}
\end{con}

Our claims regarding the bulk and defect chiral algebras  are strongly reminiscent of the AGT correspondence \cite{Alday:2009aq,Wyllard:2009hg} and of its generalization to include surface defects \cite{Alday:2010vg}. Strictly speaking, the AGT correspondence applies to the $(2,0)$ theory compactified on a complex curve $\CC$ and subjected to the $\Omega_{\epsilon_1,\epsilon_2}$ deformation in the remaining four non-compact directions. Similarly, the construction of \cite{Alday:2010vg} applies to the same compactification accompanied by a codimension two defect that also wraps $\CC$ and spans two of the four non-compact directions. In the story with no defects wrapping the UV curve, the resulting partition functions enjoy $\WW_{\gf}$ symmetry with central charge
\begin{equation}
c_{2d}=r_{\gf}+\left(b+\frac{1}{b}\right)^2d_{\gf}h_{\gf}^{\vee}~,
\end{equation}
where $b^2\colonequals\epsilon_1/\epsilon_2$. In the presence of defects wrapped on $\CC$, the partition functions display affine $\gf$ invariance at level \cite{Tachikawa:2011dz}
\begin{equation}
k_{2d}=-h^{\vee}-\frac{1}{b^2}~.
\end{equation}
Our construction, on the other hand, does not involve an explicit $\Omega$ deformation and features the six-dimensional theory in flat space. Nevertheless, at the level of chiral algebras, we reproduce these symmetries for the case $b^2=1$ (\ie, $\epsilon_1=\epsilon_2$) in the bulk case and $b^2\to\infty$ (\ie, $\epsilon_2=0$) in the presence of defects. That these particular values of $b$ should arise is somewhat reasonable. In the bulk case, $b\neq1$ would break the $\sof(4)$ symmetry of the transverse space, whereas the construction used in this paper respects that symmetry. In the defect case, $b^2\neq\infty$ would imply an effective compactification of the plane transverse to the defect. In such a scenario, one would expect to find a four-dimensional stress tensor in the resulting defect theory. In the construction considered here, such a four-dimensional stress tensor is certainly absent, and one sensibly discovers that the defect chiral algebra is at such a level that it contains no meromorphic stress tensor. The connection between the chiral algebras described in this paper and the AGT correspondence will be pursued in greater detail in future work.

\bigskip

The organization of this paper is as follows. In \S\ref{sec:chiral_derivation}, we recall the logic of \cite{Beem:2013sza} and provide the specifics of its application to the $\ospf(8^\star|4)$ superconformal algebra. We further characterize the local operators that may play a role in the protected chiral algebra of six-dimensional superconformal theories. In \S\ref{sec:abelian_case}, we consider the simple case of the abelian $(2,0)$ theory, which is a free theory and completely tractable. This serves to illustrate some aspects of the correspondence that will prove useful in the more abstract case of the interacting theories. In \S\ref{sec:nonabelian_case}, we review what is known about the half-BPS spectrum of the $(2,0)$ SCFTs and motivate the bulk chiral algebra conjecture. We show that this conjecture passes a number of checks, both at the level of the superconformal index, and at the level of three-point functions for the $A_n$ theory at large $n$. In \S\ref{sec:defects}, we address the case of half-BPS defect operators and motivate the above-stated defect chiral algebra conjecture. Various technical details and useful points of reference are included in several appendices. In particular, appendix \ref{app:characters} discusses the construction of the irreducible characters of the $\ospf(8^\star|4)$ which may be useful in future work.

\section{Chiral symmetry in a protected sector}
\label{sec:chiral_derivation}

In this section we set up the general algebraic machinery that is responsible for the existence of a protected chiral algebra of six-dimensional $(2,0)$ theories. Our approach is a direct generalization of the approach of \cite{Beem:2013sza} that was used to uncover a similar structure in four-dimensional $\NN=2$ SCFTs. That such a  generalization should exist is made apparent by observing that the four-dimensional $\NN=2$ superconformal algebra is a subalgebra of the six-dimensional $(2,0)$ superconformal algebra. The six-dimensional case turns out to be somewhat richer, however, due to the intricacies of $(2,0)$ superconformal representation theory. We will keep the exposition relatively brief. We refer the interested reader to the early sections of \cite{Beem:2013sza} for a detailed description of the analogous four-dimensional case.

We first select a fixed  \emph{chiral algebra plane} $\Rb^2\subset\Rb^6$ on which our chiral algebra will live. The first order of business is to determine the maximal subalgebra of the full superconformal algebra that fixes this plane. The $(2,0)$ superconformal algebra is reviewed in detail in Appendix \ref{app:algebras}. It is isomorphic to the $D(4,2)$ superalgebra,%
\footnote{This is the \emph{complexified} superalgebra. The relevant real form is $\ospf(8^\star|4)$. Our construction is perhaps most naturally phrased in terms of complexified algebras, but we will not be overly concerned with the distinction between complexified algebras and their real forms. Using the natural real forms can be  mnemonically helpful, \eg, in distinguishing the $\slf(2)$ M\"obius transformations from the $\suf(2)_R$ subalgebra of the $\uspf(4)$ R-symmetry that will be introduced shortly.}
the maximal bosonic subalgebra of which is the product of the $\sof(6,2)$ conformal algebra times a $\uspf(4)$ R-symmetry algebra. The choice of $\Rb^2\subset\Rb^6$ breaks $\sof(6,2)$ to the $\slf(2)\times\ol{\slf(2)}$ conformal algebra on the plane, times the $\sof(4) \cong \suf(2)_1 \times \suf(2)_2$ algebra of rotations in the transverse $\Rb^4$. We can regard $\ol{\slf(2)}\times\suf(2)_2\times\uspf(4)$ as the bosonic subalgebra of the superalgebra $D(2,2)\subset D(4,2)$, so all told we are concerned with the embeddings
\begin{equation}\label{eq:algebraembedding}
\slf(2)\times\suf_1(2)\times\left(\ol{\slf(2)} \times \suf(2)_2 \times \uspf(4) \right) \subset   \slf(2) \times \suf(2)_1 \times D(2, 2) \subset D(4, 2)~.
\end{equation}
Crucially, $D(2,2)$ contains a $\suf(1,1|2)$ subalgebra, which is a necessary condition for the cohomological construction of \cite{Beem:2013sza} to go through.\footnote{An inequivalent choice of maximal subalgebra preserving the plane is $D(2,1)\times D(2,1)\subset D(4,2)$, but this is not relevant for our purposes since $D(2,1)$ does \emph{not} have an $\suf(1,1|2)$ subalgebra.}

Let us describe this embedding more explicitly. Introducing coordinates $x_\mu$, $\mu=1,\ldots,6$ for $\Rb^6$, we take the chiral algebra plane to have complex coordinates $z=x_1+ix_2$ and $\zb=x_1-ix_2$. The generators of the two-dimensional conformal algebra $\slf(2)\times\ol{\slf(2)}$ that acts on this plane can be identified as follows (see Appendix \ref{app:algebras} for our conventions):
\begin{alignat}{6}
\label{eq:two_dimensional_conformal_generators}
&L_0	 &~=~&\tfrac12\left(\HH+\LL_1\right)~,		\qquad\quad &
&L_{+1}	 &~=~&\KK^{21}~,							\qquad\quad &
&L_{-1}	 &~=~&\PP_{12}~,\\
&\bar L_0	 &~=~&\tfrac12\left(\HH-\LL_1\right)~,		\qquad\quad &
&\bar L_{+1}&~=~&\KK^{43}~,							\qquad\quad &
&\bar L_{-1}&~=~&\PP_{34}~.\nn
\end{alignat}
We have introduced Cartan generators $\LL_{1,2,3}$ that generate rotations in the $\{x_1,x_2\}$, $\{x_3,x_4\}$, and $\{x_5,x_6\}$ planes, with eigenvalues $h_{1,2,3}$, respectively.%
\footnote{Our conventions are such that the highest- and lowest-weight components of an $\sof(6)$ vector $v_\mu$ are $v_{h.w.}=(v_1+iv_2)/\sqrt{2}$ and $v_{l.w.}=(v_1-iv_2)/\sqrt{2}$.}
The $\suf(2)_1$ and $\suf(2)_2$ subalgebras correspond to self-dual and anti-self dual rotations in the $\{x_3,x_4,x_5,x_6\}$ directions, with generators
\begin{eqnarray}\label{eq:transverse_rotation_generators}
\suf(2)_1\, & : & \qquad  \MM^1_2\,, \quad \MM^2_1\,, \quad \MM^1_1 - \MM^2_2 ~\equiv~	\LL_2 + \LL_3 ~,\\
\suf(2)_2\, & : & \qquad  \MM^3_4\,, \quad \MM^4_3\,, \quad \MM^3_3 - \MM^4_4 ~\equiv~	\LL_2 - \LL_3 ~.\nn
\end{eqnarray}
The $\uspf(4)$ generators are denoted by $\RR_{\bA\bB}$ with $\bA,\bB=1,\ldots,4$. Finally the fermionic generators of the $D(2,2)$ subalgebra comprise eight \Poincare\ supercharges $\{Q_{\bA},\wt{Q}_{\bA}\}$ and their special conformal conjugates $(S_{\bA},\wt{S}_{\bA})$, transforming under a $\uspf(4)\times\suf(2)_2$ R-symmetry. The embedding of these supercharges into the six-dimensional superalgebra is given by
\begin{equation}\label{eq:supercharge_identification}
   {Q}_{\bA}\ceq \QQ_{\bA4} ~,\qquad 
\wt{Q}_{\bA}\ceq \QQ_{\bA3} ~,\qquad 
   {S}_{\bA}\ceq \SS_{\bA}^4~,\qquad
\wt{S}_{\bA}\ceq \SS_{\bA}^3~.
\end{equation}
The identification makes it clear that $(Q_{\bA},\wt{Q}_{\bA})$  and  $(S_{\bA},\wt{S}_{\bA})$ transform as doublets of $\suf(2)_2$. 

From the $D(2,2)$ supercharges, we construct \emph{four} interesting nilpotent supercharges and their conjugates, generalizing the two that appeared in \cite{Beem:2013sza}:
\begin{alignat}{4}
\label{eq:chiral_supercharge_def}
&\qq_1&~~\ceq~~& \wt{ Q}_{\bOn}-S_{\bTh}~,\qquad\qquad&&\qq_1^\dag&~~\ceq~~& \wt{S}_{\bFo}- Q_{\bTw}~,   \\ 
&\qq_2&~~\ceq~~&  Q_{\bOn}+\wt{S}_{\bTh}~,\qquad\qquad&&\qq_2^\dag&~~\ceq~~& S_{\bFo}+\wt{ Q}_{\bTw}~,\nn\\
&\qq_3&~~\ceq~~& \wt{ Q}_{\bTw}-S_{\bFo}~,\qquad\qquad&&\qq_3^\dag&~~\ceq~~& \wt{S}_{\bTh}- Q_{\bOn}~,\nn\\
&\qq_4&~~\ceq~~&  Q_{\bTw}+\wt{S}_{\bFo}~,\qquad\qquad&&\qq_4^\dag&~~\ceq~~& S_{\bTh}+\wt{ Q}_{\bOn}~.\nn
\end{alignat}
Because the $D(2, 2)$ superalgebra is the supersymmetrization of the right-moving $\ol{\slf(2)}$ conformal algebra, all of these supercharges commute with the left-moving $\slf(2)$ generators. The key point, as in four dimensions, is to define an $R$-symmetry twist of $\ol{\slf(2)}$ that is \emph{exact} with respect to the $\qq_i$. If we take the maximal subalgebra $\suf(2)_R\times\mf{u}(1)_r\subset\uspf(4)_R$,%
\footnote{This is the subalgebra under which the $\bFi$ of $\uspf(4)$ decomposes as $\bf3_0\oplus\bf1_{+1}\oplus\bf1_{-1}$.}
then such a twisted algebra $\wh{\slf(2)}$ can be defined as the diagonal subalgebra of $\ol{\slf(2)}\times\suf(2)_R$,
\begin{equation}\label{eq:twisted_sl2_definition}
\Lh_{-1} \ceq \bar L_{-1}+\RR^-  \, , \qquad  \Lh_{0} \ceq \bar L_{0}-\RR  \, , \qquad \Lh_{+1} \ceq \bar L_{+1}-\RR^+  \, .
\end{equation}
The twisted generators occur as $\qq_i$ commutators as follows,
\begin{alignat}{5}
\label{eq:twisted_L_def}
&	2\Lh_{0}&~ = ~~&\{\qq_1,\qq_1^\dagger\}&~~=~~&\{\qq_2,\qq_2^\dagger\}&~~=~~&\{\qq_3,\qq_3^\dagger\}&~~=~~&\{\qq_4,\qq_4^\dagger\}~,\\
&	\Lh_{-1}&~ = ~~&\{\qq_1, Q_{\bFo}\}	&~~=~~&-\{\qq_2,\wt{ Q}_{\bFo}\}&~~=~~&\{\qq_3, Q_{\bTh}\}&~~=~~&-\{\qq_4,\wt{ Q}_{\bTh}\}~,\nn\\
&	\Lh_{+1}&~ = ~~&-\{\qq_1,\wt{S}_{\bTw}\}&~~=~~&\{\qq_2,S_{\bTw}\}&~~=~~&-\{\qq_3,\wt{S}	_{\bOn}\}&~~=~~&\{\qq_4,S_{\bOn}\}~.\nn
\end{alignat}
In any unitary superconformal representation one will necessarily have $\Lh_{0} \geq 0$ on any $\Lh_{0}$ eigenstate, with the equality saturated if and only if all $\qq_i$ and their conjugates $\qq_i^\dagger$ annihilate the state. Additional interesting bosonic generators are those that appear in mutual commutators of the $\qq_i$. Defining $\ZZ_{ij}\ceq\{\qq_i,\qq_j\}$, we have
\begin{eqnarray}\label{eq:Z_defs}
\ZZ_{12}&~=~&\ph{-}\ZZ_{34}~=~0~,\\
\ZZ_{13}&~=~&-\ZZ_{24}~=~\MM_3^4~,\nn\\
\ZZ_{14}&~=~&\ph{-}\LL_2-\LL_3-2r =  \MM^3_3 - \MM^4_4  - 2r  ~,\nn\\
\ZZ_{23}&~=~&\ph{-}\LL_2-\LL_3+2r =  \MM^3_3 - \MM^4_4  + 2r  ~.\nn
\end{eqnarray}

We are now in a position to define the protected chiral algebra of a six-dimensional $(2,0)$ theory. In principle, the cohomology of any one of the $\qq_i$ will have the structure of a chiral algebra. It turns out that all four $\qq_i$ define the same cohomology, and so the structure of interest is the simultaneous cohomology of all four nilpotent supercharges. Let us outline the main points of the construction.

A local operator $\OO(0)$ inserted at the origin is a ``harmonic representative'' of a $\qq_i$ cohomology class if it is annihilated by $\qq_i$ and its conjugate $\qq_i^\dagger$ (separately for each $i$), which happens if and only if it obeys $[\Lh_0,\OO (0)]=0$, \ie, if it has quantum numbers satisfying
\begin{equation}\label{eq:chirality_condition}
\frac{E-h_1}{2}- R=0~.
\end{equation}
It follows that, as stated above, the cohomology classes of all four $\qq_i$ coincide. Moreover, from Eqn. \eqref{eq:Z_defs} we can deduce that a state obeying the above condition is \emph{necessarily} invariant under $\uf(1)_r$ and $\suf(2)_2$, and so must satisfy the additional relations
\begin{equation}\label{eq:extra_quantum_number_conditions}
 r=0,\qquad h_2-h_3=0~.
\end{equation}
\emph{A priori}, $h_2 = h_3$ need not be zero, so $\qq_i$ cohomology classes are allowed to form non-trivial representations of $\suf(2)_1$.

At this point, the construction of \cite{Beem:2013sza} can be carried over \emph{verbatim}. Operators obeying the condition \eqref{eq:chirality_condition} can be translated away from the origin (within the chiral algebra plane) by means of the twisted momentum operator $\Lh_{-1}$,
\begin{equation}\label{eq:twisted_translation}
\OO(z,\zb)=e^{z L_{-1} + \zb \Lh_{-1} }\OO(0,0)e^{-z L_{-1}- \zb \Lh_{-1}}~.
\end{equation}
A local operator at the origin with $[\Lh_0,\OO(0)] = 0$ is necessarily an $\suf(2)_R$ highest weight state, carrying the maximum eigenvalue $R$ of the Cartan. Indeed, if this were not the case, states with greater values of $R$ would have negative $\Lh_0$ eigenvalue, violating unitarity. We denote the whole spin $k$ representation of $\suf(2)_R$ as $\OO^{(\II_1\cdots\II_{2k})}$, with $\II_i = 1,2$. Then the operator obeying \eqref{eq:chirality_condition} is $\OO^{11\cdots1}(0)$, and the twisted-translated operator at any other point is given by
\begin{equation}\label{displaced}
\OO(z, \zb) \colonequals u_{\II_1}(\zb)\,\cdots\,u_{\II_{2k} }(\zb) \; \OO^{(\II_1\cdots\II_{2k})} (z, \zb) \,, \qquad\quad u_{\II} (\zb) \colonequals (1, \zb) \,.
\end{equation}
By construction, such an operator is annihilated by $\qq_i$, and thanks to the second line of Eqn. \eqref{eq:twisted_L_def} its $\zb$ dependence is $\qq_i$-exact. It follows that the cohomology class of the twisted-translated operator defines a purely meromorphic operator,
\begin{equation}\label{eq:cohomology_to_chiral}
[\OO(z,\zb)]_\qq~~~\leadsto~~~\OO(z)~.
\end{equation}
Operators constructed in this manner have correlation functions that are meromorphic functions of the insertion points, and enjoy well-defined meromorphic OPEs at the level of the cohomology. These are precisely the ingredients that define a two-dimensional chiral algebra.

\subsection{Elements of the \texorpdfstring{$\qq$}{Q} cohomology}
\label{subsec:chiral_op_list}

The next step is to determine precisely which operators in a $(2,0)$ SCFT have the right properties to play a role in the protected chiral algebra. The representation theory of the $(2,0)$ superconformal algebra has been worked out in \cite{Dobrev:1985qv,Dobrev:2002dt,Bhattacharya:2008zy} and is discussed in detail in Appendix \ref{app:rep_theory}. Let us summarize the salient points here.

A generic representation is specified by a set of $\sof(6)$ Dynkin labels, $[c_1,c_2,c_3]$, a pair of $\uspf(4)_R$ Dynkin labels, $[d_1,d_2]$, and the scaling dimension, $E$, of the superconformal primary operator. In terms of the $\sof(6)$ quantum numbers $(h_1, h_2, h_3)$ and $\uspf(4)$ quantum numbers $(R,r)$ introduced above, the Dynkin labels can be written as
\begin{eqnarray}\label{eq:orthogonal_to_dynkin}
&& c_1 = h_2 - h_3~, 	\quad  c_2 = h_1 - h_2~,  \quad c_3 = h_2 + h_3~,\\
&& d_1 = R-r~, \; \;\;	\quad  d_2 = 2 r~. 
\end{eqnarray}
Shortening conditions arise when certain linear relations for these quantum numbers are satisfied. In particular, (semi-)short representations come in four series, for which the quantum numbers introduced here satisfy the following conditions,
\begin{alignat}{3}
\label{eq:semishort_series}
&\AA~&:&~ E=h_1+h_2-h_3+2R+2r+6~,\qquad	&&								\\
&\BB~&:&~ E=h_1+2R+2r+4~, 				&& h_1\geqslant h_2=h_3~,\nn 	\\
&\CC~&:&~ E=h_1+2R+2r+2~, 				&& h_1=h_2=h_3~,\nn 			\\
&\DD~&:&~ E=2R+2r~,	   					&& h_1=h_2=h_3=0~.\nn
\end{alignat}
Operators satisfying \eqref{eq:chirality_condition} only appear in a select subset of these representations. The complete list, along with the location within the full representation of the relevant operator, is determined in Appendices \ref{app:characters} and \ref{app:chiral_ops}. The results are summarized in Table \ref{Tab:Q_chiral_operators}.
\begin{table}[t]
\centering
\begin{tabular}{c|c|c|c}
Series & Primary & $\qq$-chiral & Level \\[.3ex] 
\hline
$\BB^{(\star)}$ 	 & $[c_1,c_2,0];\hfill[d_1,0]$ 	& $[c_1,c_2+2,0];\hfill[d_1+2,0]$ 	& 4 \\[.3ex]
$\BB^{\ph{(\star)}}$ & $[0,c_2,0];\hfill[d_1,0]$ 	& $[0,c_2+2,0];\hfill[d_1+2,0]$ 	& 4 \\[.3ex]
$\CC^{(\star)}$ 	 & $[c_1,0,0];\hfill[d_1,1]$ 	& $[c_1+1,1,0];\hfill[d_1+2,0]$ 	& 3 \\[.3ex]
$\CC^{(\star)}$ 	 & $[c_1,0,0];\hfill[d_1,0]$ 	& $[c_1+2,0,0];\hfill[d_1+1,0]$ 	& 2 \\[.3ex]
$\DD^{\ph{(\star)}}$ & $[0,0,0];\hfill[d_1,2]$ 		& $[0,1,0];\hfill[d_1+2,0]$ 		& 2 \\[.3ex]
$\DD^{(\star)}$ 	 & $[0,0,0];\hfill[d_1,1]$ 		& $[1,0,0];\hfill[d_1+1,0]$ 		& 1 \\[.3ex]
$\DD^{\ph{(\star)}}$ & $[0,0,0];\hfill[d_1,0]$ 		& $[0,0,0];\hfill[d_1+0,0]$ 		& 0
\end{tabular} 
\caption{Summary of superconformal representations that contain chiral algebra currents. The quantum numbers of the primary and the $\qq$-chiral operators are displayed, along with the level in the representation where one may find the $\qq$-chiral operators. Representations labelled with a star are those that seem to be absent from actual $(2,0)$ theories.\label{Tab:Q_chiral_operators}}
\end{table}

Of the representations listed, the most familiar are those in the $\DD$ series. In these representations the superconformal primary is quarter-BPS (half-BPS if $d_2=0$). It is interesting to note that in practice, all $\DD$ series multiplets in the known $(2,0)$ theories are believed to transform in representations that appear in the tensor product of sufficiently many copies of the $[1,0]$, and for this reason representations of type $\DD[0,0,0;d_1,1]$ are expected to be absent \cite{Bhattacharyya:2007sa}.

The half-BPS operators form a ring, the \emph{half-BPS ring}, which is a generalization of the chiral ring in four-dimensional supersymmetric theories. An important property of the half-BPS operators is that they are the operators with the lowest possible dimension given their $\suf(2)_R$ quantum numbers. Using this fact in conjunction with $\suf(2)$ selection rules, one quickly sees that the chiral algebra operator associated to a generator of the half-BPS ring can never appear as a normal ordered product. This means that the generators of the half-BPS ring are necessarily mapped to generators of the chiral algebra. This is a structurally identical result to the fact that in the four-dimensional case, generators of the so-called ``Hall-Littlewood chiral ring'' are mapped to chiral algebra generators.

The $\BB$ and $\CC$ series appearing in Table \ref{Tab:Q_chiral_operators} are more exotic representations that satisfy semi-shortening conditions at level two or greater. Although these are somewhat unfamiliar, we will see in the rest of this paper that the presence of $\BB$ series representations is necessary and natural from the point of view of the protected chiral algebra.

\section{The free tensor multiplet}
\label{sec:abelian_case}

Having established this basic machinery, let us consider the chiral algebra of the abelian $(2,0)$ theory. This is the theory of a free tensor multiplet, and so the chiral algebra can be constructed explicitly. The tensor multiplet lies in an ultra-short representation of type $\DD[0,0,0;1,0]$ that comprises a scalar, two Weyl fermions, and a two-form with self-dual field strength. The quantum numbers of these fields are summarized in Table \ref{Tab:free_tensor_fields}. 
\medskip
\begin{table}[h]
\centering
\begin{tabular}{c|c|c|c|c}
Operator & ~~$\Delta$~~ & ~~$\sof(6)$~~ & ~~$\uspf(4)_R$~~ & ~~$\Lh_0 (\psi_{h.w.})$~~ \\[.3ex] 
\hline
$\Phi_{\bf I}$ 			& $2$		& $\bf1$		& $\bf5$	& 	$0$		\\[.3ex]
$\lambda_{a\bA}$ 		& $\frac52$	& $\bf4$		& $\bf4$	& 	$\hf$	\\[.3ex]
$\omega^+_{(ab)}$ 		& $3$		& $\bf10$		& $\bf1$	& 	$1$		\\[.3ex]
\end{tabular} 
\caption{Field content of the abelian tensor multiplet.\label{Tab:free_tensor_fields}}
\end{table}
\medskip
Of these basic fields, the only one satisfying \eqref{eq:chirality_condition} is the $\uspf(4)$ highest-weight component of the scalar multiplet. In our conventions, this is the field
\begin{equation}\label{eq:free_scalar_hw}
\Phi_{h.w.}=\frac{\Phi_{\bOn}+i\Phi_{\bTw}}{\sqrt{2}}~.
\end{equation}
Other fields and $\uspf(4)_R$ descendants of the scalar have strictly positive eigenvalues under $\Lh_0$. The meromorphic operator associated to $\Phi_{h.w.}$ can be constructed using twisted translation in the plane as was described in \S\ref{sec:chiral_derivation}, leading to the following cohomology class,
\begin{equation}\label{eq:twisted_scalar}
\Phi(z)\colonequals \left[\tfrac{1}{\sqrt2}\left(\Phi_{\bOn}(z,\zb)+i\Phi_{\bTw}(z,\zb)\right) + \zb\Phi_{\bTh}(z,\zb)+\tfrac{\zb^2}{\sqrt2}\left(\Phi_{\bOn}(z,\zb)-i\Phi_{\bTw}(z,\zb)\right)\right]_{\qq}~.
\end{equation}
The singular part of the meromorphic $\Phi\times\Phi$ OPE follows directly from the free field OPE of the scalar fields. Specifically, if we normalize the six-dimensional operators to have canonical OPEs,
\begin{equation}\label{eq:6d_abelian_OPE}
\Phi_{\bf I}(x)\Phi_{\bf J}(y)\sim\frac{\delta_{\bf IJ}}{|x-y|^4}~,
\end{equation}
then the resulting chiral algebra OPE takes a familiar form,
\begin{equation}\label{eq:abelian_OPE}
\Phi(z)\Phi(w)\sim\frac{1}{(z-w)^2}~.
\end{equation}
This is the OPE of a $\uf(1)$ affine current,
\begin{equation}\label{eq:twisted_current}
\Phi(z) \leadsto J_{\uf(1)}(z)~.
\end{equation}

The other operators in the free theory that obey \eqref{eq:chirality_condition} are just the normal ordered products of holomorphic derivatives in the chiral algebra plane of $\Phi_{h.w.}$. These map in the obvious way to composites of the $\uf(1)$ current in the chiral algebra, \eg,
\begin{equation}\label{eq:twisted_abelian_operators}
\NO{\,(\partial_{12}^{k_1}\Phi_{h.w.})(\partial_{12}^{k_2}\Phi_{h.w.})(\partial_{12}^{k_3}\Phi_{h.w.})\,}\leadsto (\partial^{k_1}J_{\uf(1)}(\partial^{k_2}J_{\uf(1)}(\partial^{k_3}J_{\uf(1)})))~.
\end{equation}
The chiral algebra of the abelian $(2,0)$ theory is therefore precisely a $\uf(1)$ affine current algebra.

It is worthwhile to take a moment to understand the appearance of Virasoro symmetry. In four dimensions, Virasoro symmetry of the chiral algebra followed from the presence of a stress tensor in four dimensions. In six dimensions, we again find that Virasoro symmetry comes for free with the six-dimensional stress tensor multiplet. The six-dimensional stress tensor lies in a short representation of type $\DD[0,0,0;2,0]$, in which the stress tensor is a level-four descendant. The superconformal primary is a dimension four scalar transforming in the ${\bf 14}$ of $\uspf(4)$. In the case of the free theory, this primary is the symmetric traceless bilinear of scalar fields,
\begin{equation}\label{eq:stress_tensor_primary}
\OO^{(\bf14)}_{\bf IJ} \colonequals\NO{\Phi_{({\bf I}}\Phi_{{\bf J})}}~.
\end{equation}
As a half-BPS operator, the highest weight state of $\OO^{(\bf14)}_{\bf IJ}$ obeys \eqref{eq:chirality_condition}, and upon mapping to the chiral algebra this is identified with the un-normalized Sugawara operator in the $\uf(1)$ affine current algebra,
\begin{equation}\label{eq:twisted_sugawara}
[u^{\bf I}(\bar z) u^{\bf J}(\bar z)\OO^{(\bf14)}_{\bf IJ}(z,\bar z)]_\qq~\equalscolon~O^{(\bf14)}(z) \leadsto S(z)~\colonequals~(J_{\uf(1)}J_{\uf(1)})(z) ~.
\end{equation}
If we further canonically normalize this operator as $T(z) \colonequals \hf S(z)$, then direct computation leads to the standard OPE of a holomorphic stress tensor in two dimensions,
\begin{equation}\label{eq:stress_tensor_OPE}
T(z)T(0)\sim \frac{1/2}{z^4}+\frac{2T(0)}{z^2}+\frac{\partial T(0)}{z}~,
\end{equation}
where the Virasoro central charge is that of a $\uf(1)$ current algebra, namely $c_{2d}=1$. As was the case in four dimensions, we see that although the holomorphic stress tensor in the chiral algebra arises from the stress tensor multiplet in six dimensions, it corresponds to an operator in that multiplet which is \emph{not} the six-dimensional stress tensor itself.

This evaluation of the chiral algebra central charge for the abelian theory is useful since it determines for us the constant of proportionality between the two-dimensional and six-dimensional central charges. Recall that the Weyl anomaly of a $(2,0)$ theory takes the form \cite{Deser:1993yx}
\begin{equation}\label{eq:weyl_anomaly}
\AA_{6d}=aE_6+c_1I_1+c_2I_2+c_3I_3+\text{scheme dependent}~,
\end{equation}
where $E_6$ is the Euler density and $I_{1,2,3}$ are certain Weyl invariants whose precise form is unimportant for our purposes. The ratios of the two- and three-point functions of stress tensor multiplets in the $(2,0)$ theories are fixed in terms of the coefficients $c_i$ of the Weyl invariants, and these constants in turn have their ratios fixed by supersymmetry \cite{Bastianelli:2000hi}. There will therefore exist a universal constant of proportionality between $c_{2d}$ and any one of the $c_i$ that follows from supersymmetry and therefore holds for any choice of $\gf$. Having determined this constant in the abelian theory, the same result will necessarily hold for any $(2,0)$ theory. We have the general result
\begin{equation}\label{eq:central_charge_proportionality}
\frac{c_{2d}(\gf)}{c_{i}(\gf)}=\frac{1}{c^{tens}_{i}}~.
\end{equation}
This relation will prove useful in the discussion of the non-abelian theories to come. Notice that, in contrast with the four-dimensional case, the central charge of the chiral algebra of the $(2,0)$ theories is always positive.

\section{Chiral algebras of interacting \texorpdfstring{$(2,0)$}{(2,0)} theories}
\label{sec:nonabelian_case}

A direct analysis in the style of the previous section is not possible when $\gf$ is non-abelian. Nevertheless, we find compelling evidence in favor of the bulk chiral algebra conjecture put forward in the introduction. Let us first make some elementary observations that serve to motivate our claim.

Recall that the moduli space of vacua for the $(2,0)$ theory of type $\gf$ is the orbifold
\begin{equation}\label{eq:moduli_space}
\MM_{\gf}=(\Rb^5)^{r_{\gf}}/W_{\gf}~.
\end{equation}
where $r_{\gf}$ is the rank and $W_{\gf}$ the Weyl group of the Lie algebra. Let us further define the following complex subspace of $\MM_{\gf}$,
\begin{equation}\label{eq:restricted_moduli_space}
\MM^{1/2}_{\gf}\colonequals \Cb^{r_{\gf}}/W_{\gf}~.
\end{equation}
The spectrum of half-BPS operators in the $(2,0)$ theories has been studied in, \eg, \cite{Bhattacharyya:2007sa} (see also \cite{Aharony:1997th,Aharony:1997an}). These papers found confirmation of a folk theorem that states that the ring of BPS operators of an SCFT is isomorphic to the ring of holomorphic polynomials on (an appropriate subspace of the) moduli space of the theory. In the present case, this amounts to the statement that the ring of half-BPS operators in the $(2,0)$ theory of type $\gf$ is isomorphic to the holomorphic polynomial ring on $\MM^{1/2}_{\gf}$.

This ring can be given a simple description using the Harish-Chandra isomorphism. It is freely generated, with generators given by elements $\OO_i$, $i=1,\ldots, r_{\gf}$ that correspond to the Casimir invariants of $\gf$. The degree of each generator is equal to the degree of the invariant. In the language of superconformal representations, this means that the generators of the half-BPS ring live in $\DD[0,0,0;k_i,0]$ multiplets where $k_i$ is the degree of the $i$'th Casimir invariant. This was understood explicitly in \cite{Bhattacharyya:2007sa,Aharony:1997th,Aharony:1997an} for the $A_{n-1}$ theories, where this is a single Casimir invariant of degree $k=2,3,\ldots,n$. The generalization to other choices of $\gf$ is straightforward.

In \S\ref{sec:chiral_derivation} we saw that the meromorphic currents associated to generators of the half-BPS ring are necessarily generators of the associated chiral algebra. A minimal guess would then be that the chiral algebra for the $(2,0)$ theory of type $\gf$ is a $\WW$ algebra generated by precisely these currents. This guess is made more appealing upon noting that the chiral algebra $\WW_{\gf}$ that appears in the AGT correspondence \cite{Alday:2009aq,Wyllard:2009hg} for class $\SS$ theories of type $\gf$ has exactly such a structure (see, \eg, \cite{Bouwknegt:1992wg}). Indeed, our conjecture is that the protected chiral algebra of the type $\gf$ theory is precisely $\WW_{\gf}$, and we will find compelling evidence in favor of this claim.

Before moving on to specific checks, we can determine the central charge of the non-abelian chiral algebra independent of any guesswork by using Eqn. \eqref{eq:central_charge_proportionality}. The six-dimensional Weyl anomaly for $\gf=A_{n-1}$ has been determined explicitly in \cite{Tseytlin:2000sf}, and the relevant anomaly coefficients obey the following relation,
\begin{equation}\label{eq:anomaly_coefficient_An}
c_i(A_{n-1})=(4n^3-3n-1)c^{tens}_{i}~.
\end{equation}
Consequently, the central charge of the chiral algebra of the $A_{n-1}$ theory takes a suggestive form,
\begin{equation}\label{eq:2d_central_charge_An}
c_{2d}(A_{n-1})=4n^3-3n-1~.
\end{equation}
This is precisely the value of the central charge of the $A_{n-1}$ Toda CFT for $b^2=1$, or in the language of the AGT correspondence \cite{Alday:2009aq}, for $\epsilon_1=\epsilon_2$.

For the sake of completeness, we can also derive the result for the more general case. Here the anomaly coefficients take the form\footnote{Although we are not aware of this result appearing explicitly in the literature, this is the unique
expression compatible with the known central 
charge of the $A_n$ series and with the structure of $R$-symmetry anomaly polynomials, which are known for any $\gf$ \cite{Harvey:1998bx,Intriligator:2000eq}.}
\begin{equation}\label{eq:anomaly_coefficient_general}
c_i(\gf)=(4d_{\gf}h^{\vee}_{\gf}+r_{\gf})c^{tens}_{i}~,
\end{equation}
where $d_{\gf}$, $h^{\vee}_{\gf}$, and $r_{\gf}$ are the dimension, dual Coxeter number, and rank of $\gf$, respectively. The prediction for the central charge of the chiral algebra is then
\begin{equation}\label{eq:2d_central_charge_general}
c_{2d}(\gf)=4d_{\gf}h^{\vee}_{\gf}+r_{\gf}~,
\end{equation}
which again matches the relevant Toda central charge for $b^2=1$.


\subsection{Testing with the superconformal index}
\label{subsec:index_test}

Certain limits of the superconformal index for the $A_n$ theories have been computed via supersymmetric localization in five-dimensional supersymmetric Yang-Mills theory \cite{Kim:2012ava,Kim:2013nva}. The most general superconformal index (defined here with respect to the supercharge $\QQ_{\bOn4}$) takes the form \cite{Bhattacharya:2008zy}
\begin{equation}\label{eq:refined_index_def}
\II(p,q,s,t)\colonequals\Tr(-1)^Fe^{\beta\{\QQ_{\bOn4},\SS_{\bFo}^4\}}q^{E-R}p^{h_2-h_3+2r}t^{R-r}s^{h_2+h_3}~.
\end{equation}
The states that contribute to this index obey a shortening condition,
\begin{equation}\label{eq:index_shortening}
\{\QQ_{\bOn4},\SS_{\bFo}^4\}=E-2R-2r-h_1-h_2+h_3=0~.
\end{equation}
The index undergoes a radical simplification when the fugacities are specified so that the combinations of Cartan generators that appear in the exponents all commute with some additional supercharge. In particular, we may choose the fugacities so that the index has an enhanced supersymmetry with respect to $\QQ_{\bTw3}$. In this case, the resulting partition function will only receive contributions from operators that obey the additional shortening condition
\begin{equation}\label{eq:additional_index_shortening}
\{\QQ_{\bTw3},\SS_{\bTh}^3\}=E-2R+2r-h_1+h_2-h_3=0~.
\end{equation}
The relevant index with this property is equivalent to the \emph{unrefined index} studied in the aforementioned paper. In our conventions, the unrefined index is recovered by setting $t=1$, whereupon the index becomes independent of $p$ as well, and we are left with an index that depends on only two fugacities
\begin{equation}\label{eq:unrefined_index_def}
\II(q,s)\colonequals \Tr(-1)^Fe^{\beta\{\QQ_{\bOn4},\SS_{\bFo}^4\}}q^{E-R}s^{h_2+h_3}~.
\end{equation}
This is the six-dimensional analogue of the \emph{Schur index} that was defined for four-dimensional $\NN=2$ SCFTs in \cite{Gadde:2011uv}.

The operators that contribute to this index are \emph{exactly} the $\qq$-chiral operators defined in \S\ref{sec:chiral_derivation}. As a consequence, this index can be reinterpreted as a Witten index of the associated chiral algebra. Recall that the construction of \S\ref{sec:chiral_derivation} includes an $SU(2)$ global symmetry inherited from the $\suf(2)_1$ rotations transverse to the chiral algebra plane in six dimensions. In particular, the combination $h_2+h_3$ plays the role of (twice) the Cartan of this flavor symmetry, and we have
\begin{equation}\label{eq:6d_index_is_2d_index}
\II(q,s) = \II_{2d}(q,s) \colonequals \Tr_{\HH_\chi(n)}(-1)^Fq^{L_0}s^{2j_1}~,
\end{equation}
where we have denoted the Hilbert space of the chiral algebra by $\HH_\chi(n)$. In terms of the six-dimensional Lorentz group, $\qq$-chiral operators necessarily occur in representations with Dynkin labels $[c_1,c_2,0]$. The spin-statistics theorem in six dimensions implies that $c_1+c_3$ is equal to fermion number (mod $2$). Now because $c_1=2j_1$ for a $\qq$-chiral operator, it follows that $j_1$ is half-integral for fermionic operators appearing in the index and integral for bosonic operators. Consequently there can be no cancellation between bosonic and fermionic operators that contribute to the unrefined index.\footnote{This is a notable feature that, in particular, does not hold for the Schur index in four dimensions. In that setting, there can be ``accidental'' cancellations between protected operators that individually do contribute to the index.}

In \cite{Kim:2013nva}, the unrefined index of the worldvolume theory of $n$ coincident M5 branes was computed in $U(n)$ five-dimensional SYM. The resulting expression is relatively simple,
\begin{equation}\label{eq:unrefined_index_branes}
\II(q,s;n)=\prod_{k=1}^n\prod_{m=0}^\infty\frac{1}{1-q^{k+m}}=\PE\Bigg[\frac{q+q^2+\cdots+q^n}{1-q}\Bigg]~,
\end{equation}
where $\PE$ denotes plethystic exponentiation,
\begin{equation}\label{eq:plethystic_definition}
\PE\big[f(x)\big]\ceq\exp\Bigg[\sum_{m=1}^\infty\frac{f(x^m)}{m}\Bigg]~.
\end{equation}
Since the calculation was done in the $U(n)$ theory, it contains an extra factor corresponding to the index of the free tensor multiplet that describes the center of mass degrees of freedom. In other words, for the interacting theory we have
\begin{equation}\label{eq:unrefined_index_irreducible}
\II_{A_{n-1}}(q,s)=\frac{\II_{(2,0)}(q,s;n)}{\II_{(2,0)}(q,s;1)}=\prod_{k=2}^{n}\prod_{m=0}^\infty\frac{1}{1-q^{k+m}}=\PE\Bigg[\frac{q^2+\cdots+q^n}{1-q}\Bigg]~.
\end{equation}
Note that this index is actually independent of the fugacity $s$. In conjunction with the above argument for the absence of cancellation between states with the same $\suf(2)_1$ spins, this implies that in the $A_n$ series theories there are no $\qq$-chiral operators transforming in representations with non-zero $c_1$ -- namely the only superconformal representations from the list in \S\ref{sec:chiral_derivation} that actually make an appearance will be $\BB[0,c_2,0;d_1,\{0,1\}]$ and $\DD[0,0,0;d_1,\{0,2\}]$. Consequently all operators contributing to the unrefined index are bosonic, and the index can be reinterpreted as the \emph{partition function} of the chiral algebra,
\begin{equation}\label{eq:index_is_partition_function}
\II_{A_{n-1}}(q)=\Tr_{\HH_\chi(n)}q^{L_0}~.
\end{equation}

The index in Eqn. \eqref{eq:unrefined_index_irreducible} has precisely the form of the vacuum character of a chiral algebra generated by currents of spins $s=2,3,\ldots,n$ (with no null states appearing in the vacuum Verma module). The algebra $\WW_{A_{n-1}}$ is precisely such a chiral algebra, and indeed when the central charge is set as in \eqref{eq:2d_central_charge_An} then the vacuum module can be seen to contain no null states (aside from those obtained by acting with $L_{-1}$ and the higher-spin equivalents) \cite{Bouwknegt:1992wg}. In fact, given the stated spectrum of generating currents, $\WW_{A_{n-1}}$ is very likely to be the \emph{unique} solution of crossing symmetry with no nulls (up to the choice of central charge).%
\footnote{For the $A_1$, $A_2$, $A_3$, and $A_5$ cases the corresponding $\WW_{\gf}$ is the only solution to crossing symmetry with the given generators, while for $A_4$ there is an additional solution for which the vacuum module contains singular vectors for arbitrary central charge \cite{Blumenhagen:1990jv,Hornfeck:1992tm}.} 
This provides compelling support for the claim that the protected chiral algebra of a $(2,0)$ theory is the corresponding $\WW_{\gf}$ (at least for the $A_n$ series). Turning the logic around, we have a simple prediction for the generalization of \eqref{eq:unrefined_index_irreducible} for general $\gf$:
\begin{eqnarray}\label{eq:unrefined_index_prediction}
\II_{D_n}(q,s)&=&\PE\Bigg[\frac{q^n+(q^2+q^4+\cdots+q^{2n-2})}{1-q}\Bigg]~,\\
\II_{E_6}(q,s)&=&\PE\Bigg[\frac{q^2+q^5+q^6+q^8+q^9+q^{12}}{1-q}\Bigg]~,\nn\\
\II_{E_7}(q,s)&=&\PE\Bigg[\frac{q^2+q^6+q^8+q^{10}+q^{12}+q^{14}+q^{18}}{1-q}\Bigg]~,\nn\\
\II_{E_8}(q,s)&=&\PE\Bigg[\frac{q^2+q^8+q^{12}+q^{14}+q^{18}+q^{20}+q^{24}+q^{30}}{1-q}\Bigg]~.\nn
\end{eqnarray}

\subsection{Three-point couplings at large \texorpdfstring{$n$}{n}}
\label{subsec:large_N}

As a more refined check of our claim, we can compute the three-point functions of half-BPS operators for the $A_{n-1}$ series in the large $n$ limit. While there is no way aside from our chiral algebraic approach to compute half-BPS three-point functions for general $\gf$, the result at large $n$ is accessible holographically using eleven-dimensional supergravity in $AdS_7\times S^4$.\footnote{The $D_n$ case can presumably be treated similarly.} In particular, the three-point functions of ``single-trace'' half-BPS operators can be computed. The notion of a single-trace operator that is applicable here is that of generalized free field theory, since in the $(2,0)$ theories there is no obvious sense in which gauge-invariant operators are constructed from elementary matrix-valued fields. The single trace half-BPS operators are therefore the generators of the half-BPS chiral ring, which at large $n$ comprise a single scalar operator $\OO^{(k)}$ for each $k=2,3,\ldots,\infty$ with scaling dimension $\Delta=2k$. Such an operator transforms in the $k$-fold symmetric traceless tensor representation of $\sof(5)_R$. The three-point functions of these operators are required by symmetry to take the general form
\begin{equation}\label{eq:sugra_3pt_schematic}
\langle\OO^{(k_1)}_{I_1}(x_1)\OO^{(k_2)}_{I_2}(x_2)\OO^{(k_3)}_{I_3}(x_3)\rangle=\frac{C(k_1,k_2,k_3)}{x_{12}^{\Delta_{123}}x_{23}^{\Delta_{231}}x_{31}^{\Delta_{312}}}\langle\CC_{I_1}\CC_{I_2}\CC_{I_3}\rangle~,
\end{equation}
where $x_{ij}\colonequals x_i-x_j$ and $\Delta_{ijk}$\colonequals $\Delta_i+\Delta_j-\Delta_k$. The $\CC_{I_i}$ form an orthonormal basis of traceless symmetric tensors of $\sof(5)$ and $\langle\CC_{I_1}\CC_{I_2}\CC_{I_3}\rangle$ denotes the unique scalar contraction of the three tensors. For large values of $n$ these three-point couplings scale as $n^{-3/2}$, and the leading order terms have been computed in \cite{Corrado:1999pi,Bastianelli:1999en} using the supergravity description. They were found to take the form \cite{Bastianelli:1999en}
\begin{equation}\label{eq:sugra_3pt_couplings}
C(k_1,k_2,k_3)=\frac{2^{2\alpha-2}}{(\pi n)^\frac{3}{2}}\Gamma\left(\frac{k_1+k_2+k_3}{2}\right)\left(
\frac{\Gamma\left(\frac{k_{123}+1}{2}\right)\Gamma\left(\frac{k_{231}+1}{2}\right)\Gamma\left(\frac{k_{312}+1}{2}\right)}
{\sqrt{\Gamma(2k_1-1)\Gamma(2k_2-1)\Gamma(2k_3-1)}}\right)~,
\end{equation}
where $k_{ijk}\ceq k_i+k_j-k_k$ and $\alpha\ceq \frac12 (k_1+k_2+k_3)$. The operators for which this formula holds are canonically normalized, with two-point couplings given by
\begin{equation}\label{eq:sugra_2pt}
\langle\OO^{(k_i)}_{I}(x)\OO^{(k_j)}_{J}(y)\rangle=\frac{\delta^{ij}\delta_{IJ}}{|x-y|^{2k_i}}~.
\end{equation}
This is the result that should be compared to the $\WW_n$ three-point couplings in the appropriate large $n$ limit.

First, recall how the chiral algebra correlation functions are obtained from (\ref{eq:sugra_3pt_couplings}, \ref{eq:sugra_2pt}). We should replace the operators $\OO^{(k)}_{I}(x)$ by their twisted counterparts (which we shall denote $\WW^{(k)}(z)$) as in \eqref{displaced}, whereupon the resulting correlation function will be meromorphic and interpretable as a chiral algebra correlator. This amounts to a simple transformation of \eqref{eq:sugra_3pt_schematic},
\begin{equation}\label{eq:chiral_3pt}
\langle\OO^{(k_1)}_{I_1}(x_1)\OO^{(k_2)}_{I_2}(x_2)\OO^{(k_3)}_{I_3}(x_3)\rangle\Longrightarrow
\langle\WW^{(k_1)}(z_1)\WW^{(k_2)}(z_2)\WW^{(k_3)}(z_3)\rangle=\frac{C(k_1,k_2,k_3)}{z_{12}^{\hf k_{123}}z_{23}^{\hf k_{231}}z_{34}^{\hf k_{312}}}~.
\end{equation}
Making the same replacement leads to canonical normalizations for the chiral operators,
\begin{equation}\label{eq:chiral_2pt}
\langle\WW^{(k_i)}(z)\WW^{(k_j)}(w)\rangle = \frac{\delta^{k_ik_j}}{(z-w)^{2k_i}}~.
\end{equation}
Our claim is thus that the three-point couplings $C(k_1,k_2,k_3)$ will be exactly reproduced by the structure constants of the $\WW_n$ algebra (with appropriately normalization for the currents) in the double scaling limit,
\begin{equation}\label{eq:double_scaling}
n\to\infty~,\quad c_{2d}\to\infty~,\quad \frac{c_{2d}}{4n^3}\to1~.
\end{equation}

Note that because of the double scaling, the limiting $\WW$ algebra will not be the well-known $\WW_\infty$ algebra of Pope, Shen, and Romans \cite{Pope:1989ew,Pope:1989sr}. Instead to analyze this limit we will take advantage of the fact that in the limit of large central charge, the quantum chiral algebras $\WW_{\gf}$ have classical counterparts $\WW_{\gf}^{(Cl)}$. These are nonlinear Poisson algebras that can be described in terms of the Poisson brackets of a set of generators that are the classical limits of the generating currents of $\WW_{\gf}$. Moreover, the structure constants of the quantum and classical $\WW$ algebras agree at leading order in the $1/c_{2d}$ expansion.

We further make use of the fact that the Poisson algebras $\WW_n^{(Cl)}$ are limits of a one-parameter family of universal classical $\WW$ algebras $\WW_{\infty}[\mu]$ \cite{FigueroaO'Farrill:1992cv,Khesin:1993ru,Khesin:1993ww,Gaberdiel:2011wb}. For generic values of $\mu$, this algebra has one generator each of spin $2,\ldots,\infty$, while at positive integer values of $\mu$ it truncates to the $\WW_n^{(Cl)}$ algebras,
\begin{equation}\label{eq:specializing_W_mu}
\WW_n^{(Cl)}=\WW_\infty[\mu=n]~.
\end{equation}
The structure constants of $\WW_\infty[\mu]$ in a primary basis are known in closed form \cite{Campoleoni:2011hg}. We can take the double scaling limit of these results explicitly in order to determine the large $n$ correlators. The calculation itself is tedious and we found the form of the structure constants to be not illuminating in general, so we display here the appropriate double-scaled limit of some of the functions that make an appearance.

The relevant terms in the chiral algebra are the linear terms, which take the form,\footnote{Here we are using slightly different indexing conventions from those used in \cite{Campoleoni:2011hg}. For us, the indices $i,j$ are equal to the spin of the current, while in the reference the convention was that $i_{\rm there}=i_{\rm here}-1$.}
\begin{equation}\label{eq:W_current_OPE}
\WW^{(k_i)}(z)\WW^{(k_j)}(w)\sim \frac{\alpha_{k_i}(n;c_{2d})\delta^{ij}}{(z-w)^{2k_i}}+\frac{\beta_{k_ik_jk_k}(n)\WW^{(k_k)}}{(z-w)^{k_i+k_j-k_k}}+\ldots~,
\end{equation}
where in terms of the functions defined in the reference, the two-point functions are given by
\begin{equation}\label{eq:2pt_def}
\alpha_{k}(n;c_{2d})\colonequals -\frac{(2k-1)c_{2d}}{6N_{k}(n)}~,
\end{equation}
while the three-point functions take the form
\begin{equation}\label{eq:3pt_def}
\beta_{k_ik_jk_k}(n)=-\frac{(k_i+k_j-k_k-1)!C[k_i,k_j]_{\{k_k\},\{0\}}}{N_{k_i}(n)}~.
\end{equation}
In the scaling limit, the functions $N_{k_i}(n)$ are given by
\begin{equation}\label{eq:var_to_OPE_normalization}
\lim_{n\to\infty}N_{k_i}(n)=(-1)^{k_i+1}\frac{6(k_i-1)!^2}{(2k_i-1)!}n^{2k_i-4}+O(n^{2k_i-6})~.
\end{equation}
The coefficients $C[k_i,k_j]_{\{k_k\},\{0\}}$ are defined for general $n$ in \cite{Campoleoni:2011hg} and are quite complicated. In the scaling limit, the non-vanishing structure constants simplify dramatically relative to the generic case. With a significant amount of massaging they can be put into the form
\begin{equation}\label{eq:scaling_of_classical_structure}
\begin{split}
C[k_i,k_j]_{\{k_k\},\{0\}}=&
(-1)^{\frac{k_{k}+k_{i}-k_{j}-2}{2}}\times\frac{n^{k_k+k_i-k_j-2}}{2^{k_k+k_i-k_j-1}}\times
(k_{ijk}-1)!!(k_{jki}-1)!!(k_{kij}-1)!!\qquad\\
&\hspace{-.5in}\times\frac{(k_i+k_j+k_k-2)!}{(k_i+k_j+k_k-3)!!}\times
\frac{(2k_j-1)!!}{(2k_i-3)!!(2k_k-3)!!}\times
\frac{1}{(2k_k-1)(2k_j-2)!(k_{ijk}-1)!}~.
\end{split}
\end{equation}
Note that the currents for which \eqref{eq:W_current_OPE} holds differ from the canonically normalized currents with which we should compare the supergravity results. They can be rescaled to implement the canonical self-OPE according to
\begin{equation}\label{eq:rescale_currents}
\WW^{(k_i)}\rightarrow\wt{\WW}^{(k_i)}=\frac{\WW^{(k_i)}}{\sqrt{\alpha_{k_i}(n;c_{2d})}}~,
\end{equation}
which leads to the following prediction for the three-point couplings that we should be computing,
\begin{equation}\label{eq:equivalence_of_three_points}
C(k_i,k_j,k_k)=\frac{\sqrt{\alpha_{k_k}(n;c_{2d})}}{\sqrt{\alpha_{k_i}(n;c_{2d})}\sqrt{\alpha_{k_j}(n;c_{2d})}}\times \beta_{k_ik_jk_k}(n)~.
\end{equation}
Remarkably, upon plugging in the above expressions and further massaging the result, one recovers precisely the supergravity three-point functions displayed in Eqn. \eqref{eq:sugra_3pt_couplings}!\footnote{In order to recover exact agreement with \eqref{eq:sugra_3pt_couplings} it is necessary to make careful choices of the signs for the square roots appearing in \eqref{eq:equivalence_of_three_points}. With some work, these choices can be made systematically.} This agreement between $\WW$ algebra structure constants and supergravity correlation functions is an important confirmation of our claim, and goes to demonstrate the power of having identified a chiral algebraic structure in the $(2,0)$ theories. In principle, this reduces the problem of finding arbitrarily many corrections to these three-point couplings to the much better defined problem of determining the quantization of the $\WW_\infty[\mu]$ in the double scaling limit order by order in the $1/c_{2d}$ expansion.

Finally, we should remark that there is in principle some freedom in the choice of generators in terms of which one chooses to express a $\WW$ algebra, and by making a redefinition of the form $\WW^{(k_i)}\to \WW^{(k_i)}+\lambda_i\WW^{(k_j)}\WW^{(k_i-k_j)}$, for example, one may obtain an equally good set of generators. In the double scaling limit considered here, such a redefinition can be seen to only affect subleading terms as long as we are looking at \emph{non-extremal} three-point functions, \ie, $k_i+k_j < k_k$ and similarly for permutations of the indices. Strictly speaking, our check (and the results of \cite{Corrado:1999pi,Bastianelli:1999en} themselves) are only valid in the non-extremal case, due to subtleties of operator mixing in extremal correlators (\cf\ \cite{D'Hoker:1999ea}). Nevertheless, using the chiral algebra construction there is no obstruction to computing extremal three-point functions, and in fact \emph{any} extremal $n$-point function of half-BPS operators is completely determined by the corresponding chiral algebra correlation function.

\section{The chiral algebras of codimension-two defects}
\label{sec:defects}

Finally, we come to the subject of chiral algebras associated to codimension-two defects. In the theory of type $\gf$, there is an important class of half-BPS codimension-two defects labelled by embeddings $\rho:\slf(2)\to\gf$ \cite{Gaiotto:2009we}. The defect labelled by $\rho$ carries a global symmetry $\hhf \subset \gf$ that is the centralizer of the image $\rho$ in $\gf$.
  
These defects play a fundamental role in bridging six- and four-dimensional physics. Upon twisted compactification on a Riemann surface $\CC$, a $(2,0)$ theory will flow to an $\NN=2$ superconformal field theory in four dimensions. These are four-dimensional SCFTs of class $\SS$. The codimension-two defects appear in two different roles in this context: 
\begin{enumerate}
\item[(i)]If a defect fills the non-compact $\Rb^4$, and is thus located at a point on $\CC$, then its presence changes the four-dimensional theory. The resulting SCFT inherits the global symmetries of the defect.
\item[(ii)]If instead a defect wraps $\CC$ and occupies a subspace $\Rb^2\subset\Rb^4$, then it gives rise to a codimension-two defect of the four-dimensional theory.
\end{enumerate}
The four-dimensional worldvolume of such a defect enjoys $\suf(2,2|2)$ superconformal invariance, and consequently comes with a protected chiral algebra of exactly the sort discussed in \cite{Beem:2013sza}. Consider the maximal defect operator in the theory of type $\gf$ (corresponding to the trivial embedding $\rho = {\rm id}$), which carries $\gf$ global symmetry with flavor central charge given by \cite{Chacaltana:2012zy}
\begin{equation}\label{eq:defect_flavor_cc}
k_{4d}=2h^\vee~,
\end{equation}
where $h^\vee$ is the dual Coxeter number of $\gf$. Using the dictionary of \cite{Beem:2013sza}, one sees that the chiral algebra supported on this defect will necessarily include as a subalgebra an affine $\gf$ current algebra at level
\begin{equation}\label{eq:defect_flavor_cc_2d}
k_{2d}=-h^\vee~.
\end{equation}
This is the \emph{critical level}, for which the Sugawara construction becomes singular, and consequently the current algebra is without stress tensor. The most economical possibility is that the chiral algebra associated to the SCFT living on the maximal defect is \emph{just} the current algebra at the critical level. An immediate check comes from another entry in the dictionary of \cite{Beem:2013sza}: the stress tensor of the protected chiral algebra arises from the $\suf(2)_R$ symmetry current of the four-dimensional theory, which in turn belongs to the same superconformal multiplet as the four-dimensional stress tensor. The SCFT living on the defect, however, is not expected to have a local stress tensor (and hence, by supersymmetry, there should be no $R$-symmetry current). This dovetails nicely with the absence of a stress tensor in the current algebra at the critical level. Taking inspiration from the AGT correspondence and its generalization \cite{Alday:2010vg} (see also \cite{Tachikawa:2011dz,Chacaltana:2012zy}) to scenario (ii), we further conjecture that the chiral algebra associated to the defect labelled by $\rho$ is the quantum Drinfeld-Sokolov reduction of type $\rho$ of the current algebra at the critical level.

A quantitative check of our conjecture comes from an analysis of the protected spectrum of four-dimensional SCFTs of class $\SS$. By leveraging generalized S-duality and the existence of special limits where some of these theories admit Lagrangian descriptions, the general form for the superconformal index of class $\SS$ theories has been completely determined \cite{Gadde:2009kb, Gadde:2011uv,  Gaiotto:2012xa}. We may then hope to use this detailed knowledge to infer some properties of the mother $(2,0)$ theory and of its various defect operators. A salient feature of of the general formulae for the class $\SS$ index\footnote{For our purposes we should focus on the Schur limit of the index, which depends on a single superconformal fugacity $q$. According to the dictionary of \cite{Beem:2013sza}, the Schur index of the SCFT corresponds to the graded character ${\rm Tr}(-1)^Fq^{L_0}$ of the associated chiral algebra.} is the presence of factors $\hat \KK_\rho(q,\ba)$ that are naturally associated to codimension-two defects localized at punctures on $\CC$. (This is the configuration (i) mentioned above). Indeed for each puncture of type $\rho$ there is a puncture factor, $\hat \KK_\rho(q,\ba)$, which is a function of the superconformal fugacity $q$ and of the flavor fugacities $\ba$ of the global symmetry algebra $\hhf\subset\gf$. What's more, there are no other building blocks appearing in the class $\SS$ index that depend purely on local properties of the punctures. This strongly suggests that $\hat \KK_\rho(q,\ba)$ should be an intrinsic property of the codimension-two defect of type $\rho$. We may therefore suspect that $\hat \KK_\rho(q,\ba)$, once suitably normalized (see below), is the Schur index of the SCFT living on the defect of type $\rho$, which in turn is equal to the character of associated the chiral algebra. Under this assumption, our defect chiral algebra conjecture leads to a sharp prediction: $\hat \KK_\rho(q,\ba)$ must be the character of the irreducible vacuum module of the quantum Drinfeld-Sokolov reduction of the current algebra of type $\gf$ at the critical level. 

Below, we prove this statement for maximal defects in the $A_{n}$ theories. For the $D_{n}$ and $E_n$ theories a completely analogous proof is possible using the results of \cite{Lemos:2012ph,Mekareeya:2012tn}. For non-trivial embeddings, one needs to take into account the effects of the quantum Drinfeld-Sokolov reduction on the character of the chiral algebra. A proof that the resulting expressions again agree with the corresponding puncture factors can be found in \cite{forthcoming_classS}.

\subsection{The critical character}
\label{subsec:critical_character}

The irreducible vacuum character of an affine Lie algebra at the critical level has been shown to take the form \cite{Arakawa:2007aa}
\begin{equation}\label{eq:chararakawa}
\text{ch} L_\lambda = \frac{\sum_{\bar w \in \bar W}\text{sign}(\bar w)e^{\bar w(\lambda + \rho) - \rho}}{\prod_{\bar \alpha \in \bar \Delta_+}(1-q^{-\langle \lambda + \rho, \bar \alpha^\vee \rangle}) \prod_{\alpha \in \Delta_+^{\text{re}}} (1 - e^{-\alpha})}~.
\end{equation}
The notation here is standard in the mathematical literature, see \cite{Arakawa:2007aa} for a detailed explanation. Here $\lambda$ is the highest weight of the module, which we will take to be trivial, but the formula is actually valid for all cases where all but the zeroth Dynkin labels of $\lambda$ are non-negative integers.

A comparison Eqn. \eqref{eq:chararakawa} to a suitably normalized puncture factor requires some rewriting. For $\lambda = 0$ we can use the Weyl denominator formula to simplify the numerator, and we can also write out the product of the real positive roots in the denominator. Recognizing that $\chi_{\text{adj}} = \sum_{\bar \alpha \in \bar \Delta_+} (e^{\bar \alpha} + e^{-\bar \alpha}) + r$ leads to the final form,
\begin{equation}\label{eq:chararakawarewr}
\text{ch} L_0(q,{\bf a})=\PE\left(  \frac{q}{1-q} \chi_{\text{adj}}({\bf a}) - \frac{r q}{1-q} +  \sum_{\bar \alpha \in \bar \Delta_+} q^{\langle \rho, \bar \alpha^\vee \rangle} \right)~.
\end{equation}
This equation is valid for any $\gf$.

Let us now turn to the consider the factors $\hat \KK_{\rho}(q,\ba)$. In the $A_{n-1}$ theories the embeddings $\rho:\slf(2)\to\slf(n)$ are labelled by Young tableaux with $n$ boxes. We will be concerned with the trivial embedding, in which case the flavor symmetry is maximal and equal to $\slf(n)$. The corresponding Young tableau is $\Lambda = \mbox{\tiny$\yng(2)$} \ldots \mbox{\tiny$\yng(2)$}$. From Eqn. (6.9) of \cite{Gadde:2011uv}, one finds that
\begin{equation}\label{eq:hatKKunnorm}
\hat \KK_{\Lambda}(q,\ba) = \prod_{j,k=1}^{n} \PE\left( \frac{a_j a_k^{-1} q}{1-q} \right) = \PE\left( \frac{q}{1-q}  \chi_{\text{adj}}(\ba) + \frac{q}{1-q} \right)~,
\end{equation}
where the $n$ flavor fugacities $a_i$ correspond to the orthonormal basis of $A_{n-1}$ and satisfy $\prod_{i=1}^n a_i=1$. To obtain the above expression we have used the fact that the adjoint character for $A_{n-1}$ takes the form $\chi_{\text{adj}}(\ba) = \sum_{j,k = 1}^n a_j a_k^{-1} - 1$.

Before we can compare this expression to Eqn. \eqref{eq:chararakawarewr}, we should consider its normalization. Reasoning based on the superconformal index only defines $\hat \KK_{\rho}(q,\ba)$ up to an overall $q$-dependent multiplicative factor. Our interpretation suggests a natural normalization, because the index of the trivial defect (\ie, no defect whatsover) should be identically equal to $1$. The trivial defect corresponds to the principal embedding, whose associated Young tableau is the dual tableau $\Lambda^t$, so the precise version of our claim is
\begin{equation}\label{eq:precise_character_claim}
\text{ch} L_0 = \frac{\hat \KK_{\Lambda}(q,\ba)}{\hat \KK_{\Lambda^t}(q)}~.
\end{equation}
The factor for the trivial puncture, $\hat \KK_{\Lambda^t}(q)$, is again easily determined from the results of \cite{Gadde:2011uv}. We find that%
\footnote{Eqn. \eqref{eq:hatKKnorm} makes it clear that the null states that are subtracted from the Verma module are precisely those of the $W_n$ algebra. This is in fact a well-known result, for example it is an essential ingredient in the construction of Hecke eigensheaves using conformal field theory \cite{Frenkel:2005pa}.}
\begin{equation}\label{eq:hatKKnorm}
\frac{\hat \KK_{\Lambda}(q,\ba)}{\hat \KK_{\Lambda^t}(q)} = \PE\left( \frac{q}{1-q}  \chi_{\text{adj}}(\ba) - \frac{\sum_{i=2}^n q^i}{1-q} \right)~.
\end{equation}
Upon comparing \eqref{eq:hatKKnorm} and \eqref{eq:chararakawarewr} we see that the flavor fugacity dependence matches perfectly. Matching the extra $q$-dependent terms requires the relation
\begin{equation}\label{eq:simple_A_identity}
\sum_{\bar \alpha \in \bar \Delta_+} q^{\langle \rho, \bar \alpha^\vee \rangle} = \frac{n\, q - \sum_{i=1}^{n} q^i}{1-q}~,
\end{equation}
which is indeed a simple fact of life for the $A_{n-1}$ Lie algebras.


\acknowledgments
The authors have benefited from discussions with 
N.~Bobev, 
T.~Dimofte, 
M.~Douglas,
D.~Gaiotto, 
T.~Hartman, 
K.~Intriligator, 
J.~Maldacena, 
S.~Minwalla, 
S.~Razamat, 
V.~Schomerus, 
D.~Simmons-Duffin, 
and 
E.~Witten.
The authors would like to thank the Kavli Institute for Theoretical Physics for providing a hospitable and enjoyable work environment during the late stages of this project.
C. B. gratefully acknowledges support from NSF grant PHY-1314311.
L. R. gratefully acknowledges the generous support of the Simons Foundation and of the Solomon Guggenheim Foundation,
and the IAS, Princeton, for its hospitality during the initial phase of this work.
This research was also supported in part by the National Science Foundation under Grant No. NSF PHY11-25915.
\appendix

\section{The \texorpdfstring{$(2,0)$}{(2,0)} superconformal algebra}
\label{app:algebras}

\subsection{Oscillator representation}
\label{subapp:oscillator_representation}

In order to establish conventions for the the six-dimensional superconformal algebra, we utilize an oscillator representation (\cf\ \cite{Gunaydin:1985tc}). We define a set of four fermionic oscillators with their conjugates, $(c_a,\,\cd^a)$, along with four (symplectic) bosonic oscillators $\alpha_{\bf A}$. The indices $a$ and ${\bf A}$ run from one to four, and the oscillators satisfy (anti-)commutation relations
\be\label{eqn:oscillator_defs}
\{c_a,\,\cd^b\}=\delta_a^{\ph{a}b}~,\qquad [\alpha_{\bf A},\alpha_{\bf B}]=\Omega_{\bf \bA\bB}~,
\ee
where $\Omega$ is the skew-symmetric symplectic matrix with $\Omega_{\bf 14}=\Omega_{\bf 23}=1$ and other unrelated entries equal to zero.

The fermionic generators of the superconformal algebra are fermionic bilinears of the basic oscillators,
\be\label{eqn:supercharge_defs}
\QQ_{\bA a}\ceq c_a \alpha_\bA~,\qquad \SS^a_\bA\ceq \,\cd^a\alpha_\bA~,
\ee
whereas the bosonic bilinears make up the generators of the bosonic subalgebra $\mf{so}(6,2)\times \mf{usp}(4)$,
\begin{eqnarray*}
\makebox[20pt][l]{$\PP_{ab}$}		&\ceq&	c_a\,c_b~,\\
\makebox[20pt][l]{$\KK^{ab}$}		&\ceq&	\cd^a\,\cd^b~,\\
\makebox[20pt][l]{$\RR_{\bA\bB}$}	&\ceq&	\alpha_\bA\,\alpha_\bB~,\\
\makebox[20pt][l]{$\MM_a^{\ph{a}b}$}&\ceq&	c_a\,\cd^b-\tfrac14 \delta_a^{\ph{a}b}\,c_c\,\cd^c~,\\
\makebox[20pt][l]{$\HH$}	 		&\ceq&	\tfrac12\, c_a\,\cd^a~.
\end{eqnarray*}
Repeated indices are summed over.

The fermionic anti-commutators are as follows
\begin{eqnarray*}
\{\QQ_{\bA a},\QQ_{\bB b}\}&=&\Omega_{\bA\bB}\PP_{ab}~,\\
\{\SS_\bA^a,\SS_\bB^b\}&=&\Omega_{\bA\bB}\KK^{ab}~,\\
\{\QQ_{\bA a},\SS_\bB^b\}&=&\delta_a^{\ph{a}b}\RR_{\bA\bB}+\Omega_{\bA\bB}\MM_a^{\ph{a}b}+\tfrac12\delta_a^{\ph{a}b}\Omega_{\bA\bB}\HH~,
\end{eqnarray*}
while the non-vanishing commutation relations of the bosonic generators amongst themselves are given by
\begin{eqnarray*}
\lbrack\PP_{ab},\KK^{cd}\rbrack&=&\delta_b^{\ph{b}c} \MM_a^{\ph{a}d}+\delta_a^{\ph{a}d}\MM_b^{\ph{b}c}-\delta_a^c\MM_b^{\ph{b}d}-\delta_b^{\ph{b}d}\MM_a^c~,\\
\lbrack\PP_{ab},\MM_c^{\ph{c}d}\rbrack&=&\delta_a^{\ph{a}d}\PP_{bc}-\delta_b^{\ph{b}d}\PP_{ac}+\tfrac12\delta_c^{\ph{c}d}\PP_{ab}~,\\
\lbrack\KK^{ab},\MM_c^{\ph{c}d}\rbrack&=&\delta^b_c\KK^{ad}-\delta^a_c\KK^{bd}-\tfrac12\delta_c^{\ph{c}d}\KK^{ab}~,\\
\lbrack\MM_a^{\ph{a}b},\MM_c^{\ph{c}d}\rbrack&=&-\delta_a^{\ph{a}d}\MM_c^{\ph{c}b}+\delta_c^{\ph{b}b}\MM_a^{\ph{a}d}~,\\
\lbrack\HH,\PP_{ab}\rbrack&=&\PP_{ab}~,\\
\lbrack\HH,\KK^{ab}\rbrack&=&-\KK^{ab}~,\\
\lbrack\RR_{\bA\bB},\RR_{\bC\bD}\rbrack&=&\Omega_{\bA\bC}\RR_{\bB\bC}+\Omega_{\bB\bC}\RR_{\bA\bD}+\Omega_{\bA\bD}\RR_{\bB\bC}+\Omega_{\bB\bD}\RR_{\bA\bC}~.
\end{eqnarray*}
Finally, the fermionic charges have the following commutation relations with the bosonic subalgebra,
\begin{equation*}
\begin{aligned}[c]
[\PP_{ab},	\QQ_{\bC c}]	&=	0	~,\\
[\KK^{ab},	\QQ_{\bC c}]	&=	\delta_c^{\ph{c}b}\SS^{a}_\bC-\delta_c^{\ph{c}a}\SS^b_\bC~,\\
[\MM_a^{\ph{a}b},	\QQ_{\bC c}]	&=	\delta_c^{\ph{c}b}\QQ_{\bC a}-\tfrac14\delta_a^{\ph{a}b}\QQ_{\bC c}~,\\
[\HH,		\QQ_{\bC c}]	&=	\tfrac12 \QQ_{\bC c}	~,\\
[\RR_{\bA\bB},	\QQ_{\bC c}]	&=	\Omega_{\bA\bC}\QQ_{\bB c}+\Omega_{\bB\bC}\QQ_{\bA c}~,
\end{aligned}
\qquad\qquad
\begin{aligned}[c]
[\PP_{ab},	\SS^c_\bC]	&=	\delta_b^{\ph{b}c}\QQ_{\bC a}-\delta_a^c\QQ_{\bC b}~,\\
[\KK^{ab},	\SS^c_\bC]	&=	0 	~,\\
[\MM_a^{\ph{a}b},	\SS^c_\bC]	&=	-\delta_a^c\SS^b_\bC+\tfrac14\delta_a^{\ph{a}b}\SS^c_\bC ~,\\
[\HH,		\SS^c_\bC]	&=	-\tfrac12 \SS^c_\bC	~,\\
[\RR_{\bA\bB},	\SS^c_\bC]	&=	\Omega_{\bA\bC}\SS^c_{\bA}+\Omega_{\bB\bC}\SS^c_{\bA}	~.
\end{aligned}
\end{equation*}

\begin{table}[t]
\begin{centering}
\renewcommand{\arraystretch}{1.3}
\begin{tabular}{|c|c|c|c|c||c|c|}
\hline
Charge $\QQ_{{\bf A}a}$ & $h_1,h_2,h_3$ & $(j_1,j_2)$ & $R$ & $r$ &~$\mf{su}(2,2|2)$~&~~~$D(2,2)$~~~\tabularnewline
\hline
\hline
$\QQ_{{\bf1}1}$ &  $+++$ &   $(+, 0)$ & $+$  &  $+$ & $\QQ_{+}^1$ 		&	\tabularnewline
$\QQ_{{\bf2}1}$ &  $+++$ &   $(+, 0)$ & $+$  &  $-$ & 			 		&	\tabularnewline
$\QQ_{{\bf3}1}$ &  $+++$ &   $(+, 0)$ & $-$  &  $+$ & $\QQ_{+}^2$ 		&	\tabularnewline
$\QQ_{{\bf4}1}$ &  $+++$ &   $(+, 0)$ & $-$  &  $-$ & 			 		&	\tabularnewline
\hline
\hline
$\QQ_{{\bf1}2}$ &  $+--$ &   $(-, 0)$ & $+$  &  $+$ & $\QQ_{-}^1$ 		&	\tabularnewline
$\QQ_{{\bf2}2}$ &  $+--$ &   $(-, 0)$ & $+$  &  $-$ & 					&	\tabularnewline
$\QQ_{{\bf3}2}$ &  $+--$ &   $(-, 0)$ & $-$  &  $+$ & $\QQ_{-}^2$ 		&	\tabularnewline
$\QQ_{{\bf4}2}$ &  $+--$ &   $(-, 0)$ & $-$  &  $-$ & 					&	\tabularnewline
\hline
\hline
$\QQ_{{\bf1}3}$ &  $-+-$ &   $(0, +)$ & $+$  &  $+$ &  					& $Q_{\bOn}$\tabularnewline
$\QQ_{{\bf2}3}$ &  $-+-$ &   $(0, +)$ & $+$  &  $-$ &  $\Qt_{\dot+}^1$  & $Q_{\bTw}$\tabularnewline
$\QQ_{{\bf3}3}$ &  $-+-$ &   $(0, +)$ & $-$  &  $+$ &  					& $Q_{\bTh}$\tabularnewline
$\QQ_{{\bf4}3}$ &  $-+-$ &   $(0, +)$ & $-$  &  $-$ &  $\Qt_{\dot+}^2$	& $Q_{\bFo}$\tabularnewline
\hline
\hline
$\QQ_{{\bf1}4}$ &  $--+$ &   $(0, -)$ & $+$  &  $+$ &  					& $\tilde{Q}_{\bOn}$\tabularnewline
$\QQ_{{\bf2}4}$ &  $--+$ &   $(0, -)$ & $+$  &  $-$ &  $\Qt_{\dot-}^1$	& $\tilde{Q}_{\bTw}$\tabularnewline
$\QQ_{{\bf3}4}$ &  $--+$ &   $(0, -)$ & $-$  &  $+$ &  					& $\tilde{Q}_{\bTh}$\tabularnewline
$\QQ_{{\bf4}4}$ &  $--+$ &   $(0, -)$ & $-$  &  $-$ &  $\Qt_{\dot-}^2$	& $\tilde{Q}_{\bFo}$\tabularnewline
\hline
\end{tabular}
\par\end{centering}
\caption{\label{Tab:supercharges}Supercharge summary. All orthogonal basis quantum numbers have magnitude one half. The four-dimensional subalgebra acts on the $h_2$ and $h_3$ planes, while the two-dimensional chiral subalgebras act in the $h_1$ plane.}
\end{table}

\subsection{Subalgebras}

It will be convenient to explicitly define various subalgebras of $D(4,2)$. First of all, let us fix our conventions for the generators of various maximal and Cartan subalgebras of the bosonic symmetry groups.
There is a maximal subalgebra $\suf(2)_R \times \uf(1)_r \subset \mf{usp}(4)$, with generators $\{R_\pm,R\}$, and $r$ that we can take to be given by
\begin{equation}\label{eq:maximal_R_subalgebra}
R^+=\RR_{\bf 12}~,\quad R^-\ceq \RR_{\bf 34}~,\quad R\ceq -\tfrac12(\RR_{\bf 14}+\RR_{\bf 23})~,\quad r\ceq \tfrac12(\RR_{\bf 14}-\RR_{\bf 23})~.
\end{equation}
This is the subalgebra under which the $\bFi$ of $\uspf(4)$ decomposes as $\bFi\rightarrow\bTh_0\oplus\bOn_{+1}\oplus\bOn_{-1}$. The generators $R$ and $r$ define the orthogonal basis of weights for $\mf{so}(5)$, and are related to the $\mf{so}(5)$ Dynkin weights $d_1$ and $d_2$ according to
\begin{equation}
d_1=R-r~,\qquad d_2=2r~.
\end{equation}

The orthogonal basis for the Cartan subalgebra of $\mf{so}(6)$ is given by the generators of rotations in the three orthogonal planes in $\Rb^6$,
\begin{eqnarray*}
\LL_1&\ceq&\tfrac12(\MM_1^1+\MM_2^2-\MM_3^3-\MM_4^4)~,\\
\LL_2&\ceq&\tfrac12(\MM_1^1-\MM_2^2+\MM_3^3-\MM_4^4)~,\\
\LL_3&\ceq&\tfrac12(\MM_1^1-\MM_2^2-\MM_3^3+\MM_4^4)~.
\end{eqnarray*}
We denote the eigenvalues of these generators by $\LL_i|\psi\rangle=h_i|\psi\rangle$. These orthogonal basis quantum numbers are related to the Dynkin basis $[c_1,c_2,c_3]$ of $\mf{su}(4)$ according to
\begin{equation}
h_1=\frac12 c_1+ c_2+\tfrac12 c_3~,\quad h_2=\tfrac12 c_1+\tfrac12 c_3~,\quad h_3=-\tfrac12 c_1-\tfrac12 c_3~.
\end{equation}

There are a number of superconformal subalgebras of $D(4,2)$. In the text, a particularly important role is played by the maximal supersymmetrization of the algebra of anti-holomorphic M\"obius transformations in the $\{x_1,x_2\}$ plane, which is a $D(2,2)$ algebra. In addition, the four-dimensional $\NN=2$ superconformal algebra $\mf{su}(2,2|2)$ can be embedded such that the four-dimensional rotation group is $\suf(2)_1\times \suf(2)_2$ and the four-dimensional $R$-symmetry group is $\suf(2)_R\times {\rm diag}[\uf(1)_r,\uf(1)_{\LL_1}]$. The precise map between the supercharges for these two embeddings is shown in table \ref{Tab:supercharges}.

\section{Unitarity irreducible representations of \texorpdfstring{$\ospf(8^\star|4)$}{osp(8*|4)}}
\label{app:rep_theory}

We recall the classification of unitarity irreducible representations of the $\ospf(8^\star|4)$ superalgebra. These have been described in \cite{Dobrev:1985qv,Dobrev:2002dt,Bhattacharya:2008zy}. There are four linear relations at the level of quantum numbers that, if satisfied by the superconformal primary state in a representation, guarantee that the resulting representation is (semi-)short. We adopt the following notation for labelling these relations:
\begin{alignat}{3}
&\AA~&:&~ E=h_1+h_2-h_3+2R+2r+6~,\qquad&&\\
&\BB~&:&~ E=h_1+2R+2r+4~, && h_1\geqslant h_2=h_3~,\nn\\
&\CC~&:&~ E=h_1+2R+2r+2~, && h_1=h_2=h_3~,\nn\\
&\DD~&:&~ E=2R+2r~,	   && h_1=h_2=h_3=0~.\nn
\end{alignat}
Note that we are using conventions for the orthogonal Cartans such that the highest weight state of the $\bFi$ of $\uspf(4)$ has $R=1$ and $r=0$, and the highest weight state of the $\bFo$ of $\sof(6)$ has $(h_1,h_2,h_3)=(\hf,\hf,\hf)$.

For a representation in any one of the classes listed above, the structure of null states in the Verma module built on the superconformal primary depends on the $\sof(6)$ representation of that primary. Every short representation possesses a single primary null state, with the additional null states being obtained by the action of additional raising operators on the null primary. Different locations for the primary null state lead to different multiplet structures, which we summarize in table \ref{Tab:shortenings}. In all cases, when some of the $c_i$ are written, the last one is necessarily non-zero. The quantum numbers $d_{1,2}$ in all cases are only restricted to be non-negative integers. The multiplets of the type $\BB[c_1,c_2,0;0,0]$, $\CC[c_1,0,0;d_1,d_2]$ with $d_1 + d_2 \leq 1$, and $D[0,0,0;d_1,d_2]$ with $d_1 + d_2 \leq 2$ contain conserved currents or free fields. In particular, the stress tensor multiplet is $D[0,0,0;2,0]$.

\begin{table}[ht!]
\centering
\renewcommand{\arraystretch}{1.3}
\begin{tabular}{|>{$}r<{$} @{[} >{$}c<{$} @{,} >{$}c<{$} @{,} >{$}c<{$} @{;} >{$}r<{$} @{,} >{$}r<{$} @{]\,\,\qquad} | @{\qquad}  >{$}r<{$} @{} >{$}l<{$}   @{\qquad} |@{\qquad\,\,[\,} r @{,\,} r @{,\,} r @{\,;\,} r @{,\,} r @{\,]\,\,\qquad}|}
\hline
\hline
\AA & c_1 & c_2 & c_3 & d_1 & d_2 & \QQ_{{\bf1}4}  & \, \psi = 0 & 0 & 0 &-1 & 0 & 1\\
\AA & c_1 & c_2 & 0 & d_1 & d_2 & \QQ_{{\bf1}3} \QQ_{{\bf1}4}  & \, \psi = 0 & 0 & -1 & 0 & 0 & 2\\
\AA & c_1 & 0 & 0 & d_1 & d_2 & \QQ_{{\bf1}2} \QQ_{{\bf1}3} \QQ_{{\bf1}4}  & \, \psi = 0 & -1 & 0 & 0 & 0 & 3\\
\AA & 0 & 0 & 0 & d_1 & d_2 &  \QQ_{{\bf1}1}  \QQ_{{\bf1}2} \QQ_{{\bf1}3} \QQ_{{\bf1}4}  & \, \psi = 0 & 0 & 0 & 0 & 0 & 4\\
\hline
\BB & c_1 & c_2 & 0 & d_1 & d_2 & \QQ_{{\bf1}3} & \,\psi=0 & 0 & -1 & 1 & 0 & 1\\
\BB & c_1 & 0 & 0 & d_1 & d_2 & \QQ_{{\bf1}2} \QQ_{{\bf1}3} & \,\psi=0 & -1 & 0 & 1 & 0 & 2\\
\BB & 0 & 0 & 0 & d_1 & d_2 & \QQ_{{\bf1}1} \QQ_{{\bf1}2} \QQ_{{\bf1}3} & \,\psi=0 & 0 & 0 & 1 & 0 & 3\\
\hline
\CC & c_1 & 0 & 0 & d_1 & d_2 & \QQ_{{\bf1}2} & \,\psi=0 & -1 & 1 & 0 & 0 & 1\\
\CC & 0 & 0 & 0 & d_1 & d_2 & \QQ_{{\bf1}1} \QQ_{{\bf1}2} & \,\psi=0 & 0 & 1 & 0 & 0 & 2\\
\hline
\DD & 0 & 0 & 0 & d_1 & d_2 & \QQ_{{\bf1}1} & \,\psi=0 & 1 & 0 & 0 & 0 & 1\\
\hline
\hline
\end{tabular}
\caption{\label{Tab:shortenings}The primary null state for each of the shortened multiplets, expressed in terms of a combination of supercharges acting on the superconformal primary. The expression in the second column is schematic, since the actual null state  be a linear combination of this state with other descendants. The rightmost column contains the Dynkin labels corresponding to the combination of supercharges. Notice that the Lorentz indices are implicitly antisymmetrized because of the identical R symmetry indices on each supercharge.}
\end{table}

This structure of null states makes the decomposition rules for long multiplets transparent. Starting with a generic multiplet approaching the $\AA$-type bound for its dimension, the following decompositions take place (which decomposition occurs depends on the $\sof(6)$ representation of the long multiplet):
\begin{alignat}{3}
&\psi\lbrack E^*+\delta;c_1,c_2,c_3;d_1,d_2\rbrack~
&\underset{\delta\to0}{\longrightarrow}
&~~\AA\lbrack c_1,c_2,c_3;d_1,d_2\rbrack&~\oplus~&\AA\lbrack c_1,c_2,c_3-1;d_1, d_2+1\rbrack~,\\
&\psi\lbrack E^*+\delta;c_1,c_2,0;d_1,d_2\rbrack~
&\underset{\delta\to0}{\longrightarrow}
&~~\AA\lbrack c_1,c_2,0;d_1,d_2\rbrack&~\oplus~&\BB\lbrack c_1,c_2-1,0;d_1,d_1+2\rbrack~,\nn\\
&\psi\lbrack E^*+\delta;c_1,0,0;d_1,d_2\rbrack~
&\underset{\delta\to0}{\longrightarrow}
&~~\AA\lbrack c_1,0,0;d_1,d_2\rbrack&~\oplus~&\CC\lbrack c_1-1,0,0;d_1,d_2+3\rbrack~,\nn\\
&\psi\lbrack E^*+\delta;0,0,0;d_1,d_2\rbrack~
&\underset{\delta\to0}{\longrightarrow}
&~~\AA\lbrack 0,0,0;d_1,d_2\rbrack&~\oplus~&\DD\lbrack 0,0,0;d_1,d_2+4\rbrack~.\nn
\end{alignat}
There is a relatively short list of multiplets that can never appear in a recombination rule:
\begin{eqnarray}
&&\BB[c_1,c_2,0;d_1,\{0,1\}]~,\\
&&\CC[c_1,0,0;d_1,\{0,1,2\}]~,\nn\\
&&\DD[0,0,0;d_1,\{0,1,2,3\}]~.\nn
\end{eqnarray}
Amusingly, the $\qq$-chiral operators that give rise to currents of the protected chiral algebra are all selected from among these non-recombinant representations.
\section{Characters of \texorpdfstring{$\ospf(8^*|4)$}{OSp(8*|4)}}
\label{app:characters}

In this appendix we discuss a method to compute the characters for the various UIRs of $\ospf(8^*|4)$ discussed in the previous appendix. We will then use these characters to enumerate the full set of $\qq$-chiral operators given in table \ref{Tab:Q_chiral_operators} in the main text.

The characters are defined as
\[
\chi_{\RR}(\ba,\bb,q) = \text{Tr}_{\RR}(a_1^{c_1} \, a_2^{c_2} \, a_3^{c_3} \, b_1^{d_1} \, b_2^{d_2}\,  q^\Delta )
\]
where $[c_1,c_2,c_3]$ and $[d_1,d_2]$ are the $\suf(4)$ and $\sof(5)$ weights of a state in the Dynkin basis, respectively, and $\Delta$ is its scaling dimension. The trace runs over all states in the representation.

We will below write $\chi(\OO)$ to denote a monomial of the fugacities associated to an element $\OO$ of $\ospf(8^*|4)$. As an example, consider the two $\sof(5)$ raising operators $\RR^+_1$ and $\RR^+_2$ corresponding to the positive simple roots. Their respective Dynkin labels are $[2,-2]$ and $[-1,2]$ and therefore
\be
\chi(\RR^+_1) = \frac{b_1^2}{b_2^2}\,, \qquad \qquad \chi(\RR^+_2) = \frac{b_2^2}{b_1}\,.
\ee

\subsection{Long representations}
The character for a generic long representation $\LL[c_1,c_2,c_3;d_1,d_2]$, whose highest weight has scaling dimension $\Delta$, is easily constructed. It takes the form
\be \label{longchar}
\chi_{\LL} (\ba,\bb,q) =  q^\Delta \chi_{[c_1,c_2,c_3]}(\ba) \chi_{[d_1,d_2]}(\bb) P(\ba,q) Q(\ba,\bb,q) 
\ee
In this expression the terms $\chi_{[c_1,c_2,c_3]}(\bf a)$ and $\chi_{[d_1,d_2]}(\bf b)$ are just the $\suf(4)$ and $\sof(5)$ characters of the irreducible highest weight representation with the given Dynkin labels. The terms $P({\bf a},q)$ and $Q({\bf a},{\bf b},q)$ then represent the action of the supercharges and the derivatives and are defined as
\be
Q({\bf a},{\bf d},q) = \prod_{\bA, a} \left(1 + \chi(\QQ_{\bA a})\right) \qquad \qquad P(\ba,q) = \prod_{\mu = 1}^6 \left(1-\chi(\PP_{\mu})\right)^{-1}\,.
\ee

We will now rewrite equation \eqref{longchar} in a form that is useful to describe the short representations below. To this end, we notice that the characters $\chi_{[c_1,c_2,c_3]}(\bf a)$ and $\chi_{[d_1,d_2]}(\bf b)$ can be written as orbits over the Weyl group $W$,
\bea
\chi_{[c_1,c_2,c_3]}(\ba) &=& \sum_{w \in W_{\suf(4)}} w(a_1)^{c_1} w(a_2)^{c_2} w(a_3)^{c_3} M\left(w(\ba)\right) \,, \\
\chi_{[d_1,d_2]}(\bd) &=& \sum_{w \in W_{\uspf(4)}} w(b_1)^{d_1} w(b_2)^{d_2} R\left(w(\bb)\right)\,.
\eea
The factors $M(\ba)$ and $R(\bb)$ are the denominators of the Verma module characters, obtained from a product over all negative roots,
\bea
M(\ba) &=& \prod_{i = 1}^{6} \left( 1- \chi(\MM^{-}_i)\right)^{-1} = \frac{1}{\left(1-\frac{a_2}{a_1^2}\right) \left(1-\frac{1}{a_1 a_3}\right) \left(1-\frac{a_2}{a_3^2}\right) \left(1-\frac{a_1}{a_2 a_3}\right) \left(1-\frac{a_1 a_3}{a_2^2}\right) \left(1-\frac{a_3}{a_1 a_2}\right)}\,, \nonumber \\
R(\bb) &=& \prod_{j=1}^4 \left(1 - \chi(\RR^-_j)\right)^{-1} = \frac{1}{\left(1-\frac{1}{b_1}\right) \left(1-\frac{1}{b_2^2}\right) \left(1-\frac{b_1}{b_2^2}\right) \left(1-\frac{b_2^2}{b_1^2}\right)}\,.
\eea
Notice that the elements of the Weyl group act on the fugacities, whereas in the usual Weyl-Kac character formula they act on the highest weight (in a shifted way). Our expressions for the irreducible characters are however a direct rewriting of the Weyl-Kac character formula. Since the factors $P({\bf a},q)$ and $Q({\bf a},{\bf b},q)$ are invariant under the Weyl group we may also write the full character \eqref{longchar} as
\be
\begin{split}
\chi_{\LL} (\ba,\bb,q) &=  q^\Delta \sum_{w \in W}w(a_1)^{c_1} w(a_2)^{c_2} w(a_3)^{c_3} w(b_1)^{d_1} w(b_2)^{d_2} \\ & \qquad \qquad \times M(w(\ba)) R(w(\bb)) P(w(\ba),q) Q( w(\ba),w(\bb),q) 
\end{split}
\ee
where we defined $W = W_{\suf(4) \times \sof(5)}$, the Weyl group of the maximal compact bosonic subgroup of $OSp(8^*|4)$. In the following we will denote the Weyl symmetrizer sum as $\left\ldbrack \ldots \right\rdbrack_{W}$ so that we may write
\be  \label{longcharweyl}
\chi_{\LL} (\ba,\bb,q) = \left\ldbrack q^{\Delta} a_1^{c_1} a_2^{c_2} a_3^{c_3} b_1^{d_1} b_2^{d_2} M(\ba) R(\bb) P(\ba,q) Q(\ba,\bb,q) \right\rdbrack_{W}\,.
\ee
As we will shortly see, this form of the character extends most easily to short multiplets.

\subsection{Short representations}
For shortened UIRs of $\ospf(8^*|4)$ the superconformal primary state is annihilated by a subset of the supercharges\footnote{The exact null state is generically a linear combination of states obtained by acting with the supercharges and other lowering operators. This distinction is however irrelevant for the discussion in this appendix.} and, in the case of free fields or conserved currents, of the momentum operators as well. In that case there is a remarkable (but conjectural) recipe \cite{Dolan:2005wy,Bianchi:2006ti,Bhattacharya:2008zy} to compute the character: the only changes required in \eqref{longcharweyl} are to simply remove from $Q(\ba,\bb,q)$ and $P(\ba,q)$ those combinations of supercharges and momentum operators that annihilate the highest weight state, and to dial $\Delta$ to the correct scaling dimension of the superconformal primary. For example, table \ref{Tab:shortenings} shows that for a short multiplet of type $\AA[c_1,c_2,c_3;d_1,d_2]$ with $c_3 > 0$ and $d_2 > 0$ the only supercharge that generates a primary null state is $\QQ_{{\bf1}4}$ to which we associate the monomial $\chi(\QQ_{{\bf1} 4}) = b_2 q^{1/2} / a_3$. The recipe then leads to
\be
\chi_{\mathcal A[c_1,c_2,c_3;d_1,d_2]} (\ba,\bb,q) = \left\ldbrack q^{\Delta} a_1^{c_1} a_2^{c_2} a_3^{c_3} b_1^{d_1} b_2^{d_2} X(\ba) Y(\bb) P(\ba,q) Q(\ba,\bb,q) \left(1 + \frac{b_2 q^{1/2}}{a_3}\right)^{-1}\right\rdbrack_{W}
\ee
with $\Delta = 6 + c_1 / 2 + c_2 + 3 c_3 / 2 + 2 d_1 + 2 d_2$. Notice that the additional factor effectively removes from $Q(\ba,\bb,q)$ not only the primary null state but also all the states obtained from it by the action of further supercharges - this is always what we have in mind when we say that we `remove' a certain combination of supercharges.

We have implemented the recipe in \texttt{Mathematica} and obtained in this way expressions for the irreducible characters of all the shortened representations. The characters so obtained match known results and satisfy the correct recombination rules. Furthermore, when we compute the superconformal index from these characters by dialing the fugacities in an appropriate manner we find the expected form where only the ``ground states'' in the cohomology of a particular supercharge contribute. We therefore believe the resulting expressions to be correct.

Notice that the recipe requires a precise enumeration of all the different combinations of the supercharges that annihilate the superconformal primary, which is in fact rather subtle. To illustrate the general idea, consider once more the $\AA[c_1,c_2,c_3;d_1,d_2]$ multiplet with $c_3 > 0$ but now with $d_2 =0$. In that case we can act with an $\sof(5)$ lowering operator on the primary null state condition $\QQ_{{\bf1}4} \psi_{[d_1,0]} = 0$ to find that $\QQ_{{\bf2} 4}$ also annihilates the superconformal primary state,
\be
0 = \RR^-_{2} \QQ_{{\bf1}4} \psi_{[d_1,0]} = [\RR^-_{2}, \QQ_{{\bf1}4}] \psi_{[d_1,0]} = \QQ_{{\bf2} 4} \psi_{[d_1,0]}\,.
\ee
The correct character therefore becomes
\begin{multline} \label{shortcharAspecial}
\chi_{\mathcal A[c_1,c_2,c_3;d_1,0]} (\ba,\bb,q) =\\  \left\ldbrack q^{\Delta} a_1^{c_1} a_2^{c_2} a_3^{c_3} b_1^{d_1} b_2^{0} M(\ba) R(\bb) P(\ba,q) Q(\ba,\bb,q)  \left(1 + \frac{b_2 q^{1/2}}{a_3}\right)^{-1}\left(1 + \frac{b_1 q^{1/2}}{b_2 a_3}\right)^{-1}\right\rdbrack_{W}\,,
\end{multline}
where we used that $\chi(\QQ_{{\bf2} 4}) = b_1 q^{1/2}/ (b_2  a_3)$. In the remainder of this appendix we discuss how to systematically enumerate all the supercharge combinations that lead to such additional terms in the character formula.

\subsubsection{Null states and supercharge combinations}
We would like to find combinations of supercharges that annihilate the superconformal primary state, besides those obtained from the action of zero or more supercharges on the primary null states listed in table \ref{Tab:shortenings}. As in the previous example, such additional combinations arise from the action of $\suf(4)$ or $\sof(5)$ lowering operators on the null states, at least for low values of the Dynkin labels of the superconformal primary. The action of the lowering operators on the individual supercharges is as follows:
\be
\begin{tikzcd}
\arrow{rd}{\MM^-_1} \QQ_{{\bf A}1} &\\
&\arrow{ld}{\MM^-_2} \QQ_{{\bf A}2} \\
\arrow{d}{\MM^-_3} \QQ_{{\bf A}3} &\\
\QQ_{{\bf A}4} & {}
\end{tikzcd}
\qquad
\begin{tikzcd}
{} & \arrow{ld}{\RR^-_2} \QQ_{{\bf1}a} \\
\QQ_{{\bf2}a} \arrow{rd}{\RR^-_1} & {}\\
{} & \arrow{ld}{\RR^-_2} \QQ_{{\bf3}a} \\
\QQ_{{\bf4}a} & {}
\end{tikzcd} 
\ee

Let us first consider new null states appearing from the action of the Lorentz generators. Using the first diagram given above, and the specific pattern of shortening conditions in table \ref{Tab:shortenings}, we find the following rule: if 
\be
\QQ_{{\bf1} a_1} \ldots \QQ_{{\bf1} a_k} \psi_{[c_1,c_2,c_3;d_1,d_2]} = 0
\ee
then the additional shortenings with the same R symmetry indices are obtained by taking all possible actions of the lowering operators $\MM^-_i$. Therefore if
\be \label{lorentzlowerqs}
[\MM^-_{i_1}, \ldots ,[\MM^-_{i_m} ,\QQ_{{\bf1} a_1} \ldots \QQ_{{\bf1} a_k} ]] =: \QQ_{{\bf1} b_1} \ldots \QQ_{{\bf1} b_k} \neq 0
\ee
then the resulting combination of supercharges annihilates the superconformal primary and so the corresponding term needs to be subtracted from $Q(\ba,\bb,q)$ in the character formula. For example, consider the $\BB[c_1,0,0;d_1,d_2]$ multiplet. Table \ref{Tab:shortenings} shows that the primary null state is given by
\be
\QQ_{{\bf1} 2} \QQ_{{\bf1} 3} \psi_{[c_1,0,0]}  = 0
\ee
Acting with $\MM^-_3$ and then further with $\MM^-_2$, both of which annihilate the superconformal primary, we find the additional null states
\be
\QQ_{{\bf1} 2} \QQ_{{\bf1} 4} \psi_{[c_1,0,0]} = 0 \qquad \QQ_{{\bf1} 3} \QQ_{{\bf1} 4} \psi_{[c_1,0,0]} = 0 
\ee
These combinations therefore also need to be removed from the factor $Q(\ba,\bb,q)$. The explanation behind our general rule tracks the logic of this example: from a direct analysis one finds that if \eqref{lorentzlowerqs} is non-zero for a specific combination of the $\MM^-_i$ then each $\MM^-_i$ in this combination annihilates the superconformal primary and the result follows.

Next we consider the R symmetry quantum numbers. By carefully matching how many states should be removed at a given level against the number of states available one uncovers the following slightly more involved pattern. First of all, we find that for $d_2 > 3$ the above analysis suffices and there are no additional terms that need to be removed from $Q(\ba,\bb,q)$. To see what happens for $d_2 \leq 3$ let us consider the type $\DD[0,0,0;d_1,3]$ multiplet as an example. We obtain from table \ref{Tab:shortenings} that the superconformal primary is killed by $\QQ_{{\bf1}1}$ and, in agreement with the rule given above, we find three more shortenings by acting with $\MM^-_a$. Altogether this leads to
\be \label{Dexshortening1}
\QQ_{{\bf1}1} \psi_{[0,0,0;d_1,3]} = 0\,, \qquad \QQ_{{\bf1}2} \psi_{[0,0,0;d_1,3]} = 0\,, \qquad \QQ_{{\bf1}3} \psi_{[0,0,0;d_1,3]} = 0\,, \qquad \QQ_{{\bf1}4} \psi_{[0,0,0;d_1,3]} = 0\,.
\ee
Let us now demonstrate that the highest weight state in this multiplet satisfies the additional relation:
\be \label{Dexshortening2}
\QQ_{{\bf2}1}\QQ_{{\bf2}2}\QQ_{{\bf2}3}\QQ_{{\bf2}4}\psi = 0
\ee
and therefore that this combination of supercharges also needs to be removed from the character formula. In order to show \eqref{Dexshortening2} it suffices to realize that one may rewrite this expression as a linear combination of the following terms:
\be
\begin{split}
(\RR^-_2)^4 \QQ_{{\bf1}1}\QQ_{{\bf1}2}\QQ_{{\bf1}3}\QQ_{{\bf1}4}\psi&\\
(\RR^-_2)^3 [\RR^-_2,\QQ_{{\bf1}1}\QQ_{{\bf1}2}\QQ_{{\bf1}3}\QQ_{{\bf1}4}]\psi&\\
(\RR^-_2)^2 [\RR^-_2, [\RR^-_2,\QQ_{{\bf1}1}\QQ_{{\bf1}2}\QQ_{{\bf1}3}\QQ_{{\bf1}4}]]\psi&\\
\RR^-_2 [\RR^-_2 [\RR^-_2, [\RR^-_2,\QQ_{{\bf1}1}\QQ_{{\bf1}2}\QQ_{{\bf1}3}\QQ_{{\bf1}4}]]]\psi&\\
\QQ_{{\bf1}1}\QQ_{{\bf1}2}\QQ_{{\bf1}3}\QQ_{{\bf1}4} (\RR^-_2)^4 \psi&
\end{split}
\ee
with coefficients that are easily determined but unimportant for our analysis. Now, each of the states listed above is null: the first four because the commutators evaluate to a term that vanishes due to the shortening \eqref{Dexshortening1}, and the last one because $\psi$ has $d_2 = 3$.

Notice that \eqref{Dexshortening2} is precisely the shortening condition of $\AA[0,0,0;d_1,d_2]$ type acted upon with $(\RR^-_2)^4$. This is indicative of the following general pattern. Let us enumerate the shortening types by an integer $\XX = \{1,2,3,4\}$ for $\{\AA,\BB,\CC,\DD\}$, respectively. Then, in a short multiplet of type $\XX[c_1,c_2,c_3;d_1,d_2]$ with $\XX - d_2 > 0$ and $d_1 > 3$, we need to additionally remove from $Q(\ba,\bb,q)$ precisely those combinations of the supercharges that one obtains from the action of $(\RR^-_2)^{d_2 + 1}, (\RR^-_2)^{d_2 + 2}, \ldots ,(\RR^-_2)^{4}$ on the combination states of type $(\XX - d_2)[c_1,c_2,c_3;d_1,d_2]$.\footnote{Notice that the Lorentz indices in the shortening conditions are always antisymmetrized so one may take the $\sof(5)$ indices to be symmetrized. The precise statement is that one has to remove precisely one term in the sum that symmetrizes the $\sof(5)$ indices, but it does not matter which term.}

In a similar vein one finds that further factors may need to be removed if, in addition to $d_2 \leq 3$, the superconformal primary has $d_1 + d_2 \leq 3$. In that case one should also remove the supercharge combinations obtained from the action of $(\RR^-_1)^{d_1 + 1}$ on the shortening conditions of type $(\XX - d_1 - d_2)[c_1,c_2,c_3;d_1,d_2]$ obtained using the above procedure for $d_2 < 3$, as well as all the supercharge combinations obtained from all the other $\sof(5)$ lowering operators acting on this combination.

Finally, when the multiplet contains conserved currents one should further remove the action of certain momentum operators from $P(\ba,q)$. We have not implemented this in detail, relying instead on the known form of short representations of the conformal algebra \cite{Dolan:2005wy} to obtain expressions that match our expectations.

\subsubsection{Contribution to the unrefined superconformal index}
The explicit form of the characters is obviously rather involved. It does not seem wortwhile to reproduce them here, but they are available from the authors upon request. On the other hand, we can use these characters to compute the contribution to the superconformal index of each shortened multiplet. If we in addition take the unrefined limit of the superconformal index as described in section \ref{subsec:chiral_op_list} then the expressions simplify drastically. We find a non-zero contribution only for the following six cases:
\begin{align}
\BB[c_1,c_2,0;d_1,0] &: \frac{q^{4 + d_1 + c_1 / 2 + c_2 }}{1-q} \chi_{c_1}(s) & \DD[0,0,0;d_1,0] &: \frac{q^{d_1}}{1-q} \nonumber\\
\CC[c_1,0,0;d_1,0] &: \frac{q^{2 + d_1 + c_1 / 2}}{1-q} \chi_{c_1 + 2}(s) & \DD[0,0,0;d_1,1] &: \frac{q^{3/2 + d_1}}{1-q} \chi_1(s)\\
\CC[c_1,0,0;d_1,1] &: \frac{q^{7/2 + d_1 + c_1 / 2}}{1-q} \chi_{c_1 + 1}(s) & \DD[0,0,0;d_1,2] &: \frac{q^{3 + d_1}}{1-q} \nonumber
\end{align}
with
\be
\chi_{\lambda}(s) = \frac{s^{\lambda + 1}- s^{-\lambda - 1} }{s-s^{-1}}
\ee
the $\suf(2)$ character corresponding to the irrep with highest weight $\lambda$. These are the $\qq$-chiral operators described in the main text.

\section{\texorpdfstring{$\qq$}{Q}-chiral operators}\label{app:chiral_ops}

In this appendix we provide an alternative and more direct derivation of the $\qq$-chiral operators. The derivation presented below does not require the computation of characters and may give a more intuitive picture behind the presence of $\qq$-chiral operators in shortened UIRs. The results that we obtain are in complete agreement with those of the previous appendix.

As explained in the main text, a $\qq$-chiral operator satisfies the following defining set of conditions for its quantum numbers:
\be
[\hat L_0,\OO(0)]=0 \quad \Longrightarrow \quad \frac{E-h_1}{2}-R=0~,
\ee
By unitary, an operator satisfying this condition will necessarily obey the additional relations
\be
h_2=h_3~,\qquad r=0~.
\ee

Now the important question for us is where such states may appear in a UIR of the six-dimensional superconformal algebra. Let us start by determining some general properties regarding the placement of such an state in a six-dimensional represenation. First, we can see that such a state must be in the highest weight state of its $\sof(5)$ representation. Indeed, consider a state $|\psi\rangle_{[d_1,0]}$ that obeys the $\qq$-chirality conditions and has Dynkin weights $[d_1,d_2]$. The action of the positive simple roots of $\uspf(4)$ on this state will be as follows,
\begin{eqnarray*}
\RR^+_1|\psi\rangle_{[d_1,d_2]} &=& \lambda_1|\psi\rangle_{[d_1+2,d_2 -2]}~,\\
\RR^+_2|\psi\rangle_{[d_1,d_2]}  &=& \lambda_2|\psi\rangle_{[d_1-1,d_2 + 2]}~,
\end{eqnarray*}
for some coefficients $\lambda_{1,2}$. If either $\lambda$ were non-zero, one can easily see that the resulting state would violate unitarity in the sense that sums of squares of some supercharges would have negative eigenvalue when acting on that state. Consequently, a $\qq$-chiral state must necessarily be a $\sof(5)$ highest weight.

The story of $\suf(4)$ representations is not quite as simple. This is because the supercharges involved in these arguments all commute with the subgroup $\suf(2)_1\subset \suf(4)$. If we consider the action of the positive simple roots of $\suf(4)$ acting on a $\qq$-chiral state $\psi_{[c_1,c_2,c_3]}$,
\begin{eqnarray*}
\MM^+_1\psi_{[c_1,c_2,c_3]}	&\sim&	\psi_{[c_1+2,c_2-1,c_3]}~,\\
\MM^+_2\psi_{[c_1,c_2,c_3]}	&\sim&	\psi_{[c_1-1,c_2+2,c_3 -1]}~,\\
\MM^+_3\psi_{[c_1,c_2,c_3]}	&\sim&	\psi_{[c_1  ,c_2-1,c_3+ 2]}~.
\end{eqnarray*}
The second and third of these states will violate unitarity if non-zero. The first, on the other hand, could be an allowed state that also satisfies the $\qq$-chirality conditions, and indeed the first represents the action of the raising operator of $\suf(2)_1$. We may conclude that within a given representation of $\suf(4)\times \sof(5)$, the only potential $\qq$-chiral operators are of the form
$$
(\MM^{-}_1)^k|\psi\rangle_{h.w.}~,
$$
which fill out the representation of $\suf(2)_1$ in which the $\suf(4)$ highest weight transforms. For the purpose of identifying representations in which $\qq$-chiral operators reside, it will therefore be sufficient to look for highest weight states of $\suf(4)\times \sof(5)$ that are $\qq$-chiral, and subsequently include any additional states in the relevant $\suf(2)_1$ multiplet.

We can do this as follows. 
The highest weight state of any $\suf(4)\times \sof(5)$ representation appearing in the superconformal multiplet will be a linear combination of states, at least one of which will take the form of up to sixteen supercharges acting on the superconformal highest weight state:
$$
\psi_{h.w.}=\QQ\cdots\QQ \psi_{s.c.h.w.}+\ldots
$$
Thus, we can search for $\qq$-chiral operators searching of states of this form with the correct quantum numbers. 
The possible operators of this type are immediately restricted by the fact that $\hat L_0$ must have positive eigenvalues on any physical state, and there are only four supercharges whose action reduces the value of $\hat L_0$.
Thus we are actually only interested in operators of the form
$$
\QQ_{{\bf1} 1}^{n_1}\QQ_{{\bf1} 2}^{n_2}\QQ_{{\bf2} 1}^{n_3}\QQ_{{\bf2} 2}^{n_4}\psi~,
$$
with $n_i=0,1$.\footnote{There are also supercharges that do not shift the value of $\hat L_0$, and one may at first think that those could be included in the action as well. However, those supercharges will necessarily shift the value of $r$, meaning that if a $\qq$-chiral operator existed that included an action of such a supercharge, there would be another operator present with $\hat L_0=0$ and $r\neq0$, which would violate unitarity.}

The most that the $\hat L_0$ eigenvalue of any superconformal primary state can be lowered before reaching a $\qq$-chiral operator is therefore two. 
Consequently, the types of multiplets that may conceivably contain $\qq$-chiral operators are those which the superconformal primary has $\hat L_0$ eigenvalue less than or equal to two, along with some additional $r$ symmetry constraints. 
The possible cases are easily enumerated to be the following:
$$
\BB[c_1,c_2,0;d_1,0]~,\quad \CC[c_1,0,0;d_1,\{0,1\}]~,\quad \DD[0,0,0;d_1,\{0,1,2\}]~.
$$
Let us consider these options in order. 

\begin{enumerate}
\item[i] A multiplet of type $\BB[c_1,c_2,0;d_1,0]$. In this case there is a potential $\qq$-chiral operator including a term of the form $\QQ_{{\bf1} 2}\QQ_{{\bf2} 2}\QQ_{{\bf2} 1}\QQ_{{\bf2} 2}\psi_{s.c.h.w.}$.
Indeed, one can check that such a state does exist (it is not excluded by the shortening conditions), and it is in the highest weight state of the following projection:
$$
\restr{\QQ^{\otimes 4}\psi}{[c_1,c_2+2,0;d_1+2,0]}~.
$$
Note that generally speaking $c_1$ may be non-zero, in which case such a $\qq$-chiral operator will lie in an $SU(2)$ multiplet of $\qq$-chiral operators.

\item[ii(a)] The next consideration is $\CC[c_1,0,0;d_1,0]$. For such a multiplet, there are potential $\qq$-chiral operators including terms of the following forms: $\QQ_{{\bf1} 1}\QQ_{12}\psi_{scp}$, $\QQ_{{\bf1} 1}\QQ_{{\bf2} 2}\psi_{scp}$, $\QQ_{{\bf2} 1}\QQ_{{\bf1} 2}\psi_{scp}$, or $\QQ_{{\bf2} 1}\QQ_{{\bf2} 2}\psi_{scp}$. It turns out that only the first of these actually appears in the highest weight component of a $\qq$-chiral operator, which is as follows:
$$
\restr{\left(\QQ^{\otimes2}\otimes\psi\right)}{[c_1+2,0,0;d_1+1,0]}~.
$$
There will always be a non-trivial $SU(2)$ multiplet of $\qq$-chiral operators in this case.

\item[ii(b)] We also consider the case $\CC[c_1,0,0;d_1,1]$. In this case the potential $\qq$-chiral operators include terms of the form $\QQ_{{\bf2} 1}\QQ_{{\bf1} 2}\QQ_{{\bf2} 2}\psi_{scp}$ and $\QQ_{{\bf1} 1}\QQ_{{\bf1} 2}\QQ_{{\bf2} 2}\psi_{scp}$.
Again, only the first of these appears in a highest weight $\qq$-chiral operator, which is as follows:
$$
\restr{\left(\QQ^{\otimes3}\otimes\psi_{[c_1,0,0;d_1,1]}\right)}{[c_1+1,1,0;d_1+2,0]}~,
$$
We have a non-trivial $SU(2)$ multiplet again.

\item[iii(a)] Finally, we consider the (at least) quarter BPS states of type $\DD$. For $\DD[0,0,0;d_1,0]$, these are actually half BPS states, and the superconformal primary itself is $\qq$-chiral,
$$
\psi_{[0,0,0;d_1,0]}~.
$$

\item[iii(b)] For $\DD[0,0,0;d_1,1]$, the possible $\qq$-chiral states include terms of the form $\QQ_{{\bf1} 2}\psi_{scp}$ and $\QQ_{{\bf2} 2}\psi_{scp}$. The second of these is in the same multiplet as the first, which gives us a $\qq$-chiral highest weight state in the following projection:
$$
\restr{\left(\QQ\otimes\psi_{[0,0,0;d_1,1]}\right)}{[1,0,0;d_1+1,0]}~.
$$

\item[iii(c)] Finally, for $\DD[0,0,0;d_1,2]$, the only possible $\qq$-chiral states include the term $\QQ_{{\bf1} 2}\QQ_{{\bf2} 2}\psi_{scp}$, which occurs in the following projection:
$$
\restr{\left(\QQ^{\otimes2}\otimes\psi_{[0,0,0;d_1,2]}\right)}{[0,1,0;d_1+2,0]}~.
$$
\end{enumerate}

\bibliographystyle{./aux/JHEP}
\bibliography{6d_chiral}

\providecommand{\href}[2]{#2}\begingroup\raggedright\begin{thebibliography}{10}

\bibitem{Beem:2013sza}
C.~Beem, M.~Lemos, P.~Liendo, W.~Peelaers, L.~Rastelli, {\em et.~al.}, {\it
  {Infinite Chiral Symmetry in Four Dimensions}},
  \href{http://xxx.lanl.gov/abs/1312.5344}{{\tt arXiv:1312.5344}}.

\bibitem{Kac:1977qb}
V.~Kac, {\it {A Sketch of Lie Superalgebra Theory}},  {\em Commun.Math.Phys.}
  {\bf 53} (1977) 31--64.

\bibitem{Nahm:1977tg}
W.~Nahm, {\it {Supersymmetries and their Representations}},  {\em Nucl.Phys.}
  {\bf B135} (1978) 149.

\bibitem{Corrado:1999pi}
R.~Corrado, B.~Florea, and R.~McNees, {\it {Correlation functions of operators
  and Wilson surfaces in the d = 6, (0,2) theory in the large N limit}},  {\em
  Phys.Rev.} {\bf D60} (1999) 085011,
  [\href{http://xxx.lanl.gov/abs/hep-th/9902153}{{\tt hep-th/9902153}}].

\bibitem{Bastianelli:1999en}
F.~Bastianelli and R.~Zucchini, {\it {Three point functions of chiral primary
  operators in d = 3, N=8 and d = 6, N=(2,0) SCFT at large N}},  {\em
  Phys.Lett.} {\bf B467} (1999) 61--66,
  [\href{http://xxx.lanl.gov/abs/hep-th/9907047}{{\tt hep-th/9907047}}].

\bibitem{Alday:2009aq}
L.~F. Alday, D.~Gaiotto, and Y.~Tachikawa, {\it {Liouville Correlation
  Functions from Four-dimensional Gauge Theories}},  {\em Lett.Math.Phys.} {\bf
  91} (2010) 167--197, [\href{http://xxx.lanl.gov/abs/0906.3219}{{\tt
  arXiv:0906.3219}}].

\bibitem{Wyllard:2009hg}
N.~Wyllard, {\it {A(N-1) conformal Toda field theory correlation functions from
  conformal N = 2 SU(N) quiver gauge theories}},  {\em JHEP} {\bf 0911} (2009)
  002, [\href{http://xxx.lanl.gov/abs/0907.2189}{{\tt arXiv:0907.2189}}].

\bibitem{Gaiotto:2009we}
D.~Gaiotto, {\it {N=2 dualities}},  {\em JHEP} {\bf 1208} (2012) 034,
  [\href{http://xxx.lanl.gov/abs/0904.2715}{{\tt arXiv:0904.2715}}].

\bibitem{Aharony:1997th}
O.~Aharony, M.~Berkooz, S.~Kachru, N.~Seiberg, and E.~Silverstein, {\it {Matrix
  description of interacting theories in six-dimensions}},  {\em
  Adv.Theor.Math.Phys.} {\bf 1} (1998) 148--157,
  [\href{http://xxx.lanl.gov/abs/hep-th/9707079}{{\tt hep-th/9707079}}].

\bibitem{Bhattacharyya:2007sa}
S.~Bhattacharyya and S.~Minwalla, {\it {Supersymmetric states in M5/M2 CFTs}},
  {\em JHEP} {\bf 0712} (2007) 004,
  [\href{http://xxx.lanl.gov/abs/hep-th/0702069}{{\tt hep-th/0702069}}].

\bibitem{Kim:2012ava}
H.-C. Kim and S.~Kim, {\it {M5-branes from gauge theories on the 5-sphere}},
  {\em JHEP} {\bf 1305} (2013) 144,
  [\href{http://xxx.lanl.gov/abs/1206.6339}{{\tt arXiv:1206.6339}}].

\bibitem{Kim:2013nva}
H.-C. Kim, S.~Kim, S.-S. Kim, and K.~Lee, {\it {The general M5-brane
  superconformal index}},  \href{http://xxx.lanl.gov/abs/1307.7660}{{\tt
  arXiv:1307.7660}}.

\bibitem{Blumenhagen:1990jv}
R.~Blumenhagen, M.~Flohr, A.~Kliem, W.~Nahm, A.~Recknagel, {\em et.~al.}, {\it
  {W algebras with two and three generators}},  {\em Nucl.Phys.} {\bf B361}
  (1991) 255--289.

\bibitem{Hornfeck:1992tm}
K.~Hornfeck, {\it {W algebras with set of primary fields of dimensions (3, 4,
  5) and (3, 4, 5, 6)}},  {\em Nucl.Phys.} {\bf B407} (1993) 237--246,
  [\href{http://xxx.lanl.gov/abs/hep-th/9212104}{{\tt hep-th/9212104}}].

\bibitem{Rattazzi:2008pe}
R.~Rattazzi, V.~S. Rychkov, E.~Tonni, and A.~Vichi, {\it {Bounding scalar
  operator dimensions in 4D CFT}},  {\em JHEP} {\bf 0812} (2008) 031,
  [\href{http://xxx.lanl.gov/abs/0807.0004}{{\tt arXiv:0807.0004}}].

\bibitem{Beem:2013qxa}
C.~Beem, L.~Rastelli, and B.~C. van Rees, {\it {The N=4 Superconformal
  Bootstrap}},  {\em Phys.Rev.Lett.} {\bf 111} (2013) 071601,
  [\href{http://xxx.lanl.gov/abs/1304.1803}{{\tt arXiv:1304.1803}}].

\bibitem{forthcoming_20}
C.~Beem, M.~Lemos, L.~Rastelli, and B.~C. van Rees, {\it {Work in Progress}}, .

\bibitem{Freedman:1998tz}
D.~Z. Freedman, S.~D. Mathur, A.~Matusis, and L.~Rastelli, {\it {Correlation
  functions in the CFT(d) / AdS(d+1) correspondence}},  {\em Nucl.Phys.} {\bf
  B546} (1999) 96--118, [\href{http://xxx.lanl.gov/abs/hep-th/9804058}{{\tt
  hep-th/9804058}}].

\bibitem{Lee:1998bxa}
S.~Lee, S.~Minwalla, M.~Rangamani, and N.~Seiberg, {\it {Three point functions
  of chiral operators in D = 4, N=4 SYM at large N}},  {\em
  Adv.Theor.Math.Phys.} {\bf 2} (1998) 697--718,
  [\href{http://xxx.lanl.gov/abs/hep-th/9806074}{{\tt hep-th/9806074}}].

\bibitem{Baggio:2012rr}
M.~Baggio, J.~de~Boer, and K.~Papadodimas, {\it {A non-renormalization theorem
  for chiral primary 3-point functions}},  {\em JHEP} {\bf 1207} (2012) 137,
  [\href{http://xxx.lanl.gov/abs/1203.1036}{{\tt arXiv:1203.1036}}].

\bibitem{Chacaltana:2012zy}
O.~Chacaltana, J.~Distler, and Y.~Tachikawa, {\it {Nilpotent orbits and
  codimension-two defects of 6d N=(2,0) theories}},  {\em Int.J.Mod.Phys.} {\bf
  A28} (2013) 1340006, [\href{http://xxx.lanl.gov/abs/1203.2930}{{\tt
  arXiv:1203.2930}}].

\bibitem{Frenkel:2004jn}
E.~Frenkel and D.~Ben-Zvi, {\em Vertex algebras and algebraic curves}.
\newblock American Mathematical Society, 2001.

\bibitem{Frenkel:2005pa}
E.~Frenkel, {\it {Lectures on the Langlands program and conformal field
  theory}},  \href{http://xxx.lanl.gov/abs/hep-th/0512172}{{\tt
  hep-th/0512172}}.

\bibitem{Alday:2010vg}
L.~F. Alday and Y.~Tachikawa, {\it {Affine SL(2) conformal blocks from 4d gauge
  theories}},  {\em Lett.Math.Phys.} {\bf 94} (2010) 87--114,
  [\href{http://xxx.lanl.gov/abs/1005.4469}{{\tt arXiv:1005.4469}}].

\bibitem{Tachikawa:2011dz}
Y.~Tachikawa, {\it {On W-algebras and the symmetries of defects of 6d N=(2,0)
  theory}},  {\em JHEP} {\bf 1103} (2011) 043,
  [\href{http://xxx.lanl.gov/abs/1102.0076}{{\tt arXiv:1102.0076}}].

\bibitem{Dobrev:1985qv}
V.~Dobrev and V.~Petkova, {\it {All Positive Energy Unitary Irreducible
  Representations of Extended Conformal Supersymmetry}},  {\em Phys.Lett.} {\bf
  B162} (1985) 127--132.

\bibitem{Dobrev:2002dt}
V.~Dobrev, {\it {Positive energy unitary irreducible representations of D = 6
  conformal supersymmetry}},  {\em J.Phys.} {\bf A35} (2002) 7079--7100,
  [\href{http://xxx.lanl.gov/abs/hep-th/0201076}{{\tt hep-th/0201076}}].

\bibitem{Bhattacharya:2008zy}
J.~Bhattacharya, S.~Bhattacharyya, S.~Minwalla, and S.~Raju, {\it {Indices for
  Superconformal Field Theories in 3,5 and 6 Dimensions}},  {\em JHEP} {\bf
  0802} (2008) 064, [\href{http://xxx.lanl.gov/abs/0801.1435}{{\tt
  arXiv:0801.1435}}].

\bibitem{Deser:1993yx}
S.~Deser and A.~Schwimmer, {\it {Geometric classification of conformal
  anomalies in arbitrary dimensions}},  {\em Phys.Lett.} {\bf B309} (1993)
  279--284, [\href{http://xxx.lanl.gov/abs/hep-th/9302047}{{\tt
  hep-th/9302047}}].

\bibitem{Bastianelli:2000hi}
F.~Bastianelli, S.~Frolov, and A.~A. Tseytlin, {\it {Conformal anomaly of (2,0)
  tensor multiplet in six-dimensions and AdS / CFT correspondence}},  {\em
  JHEP} {\bf 0002} (2000) 013,
  [\href{http://xxx.lanl.gov/abs/hep-th/0001041}{{\tt hep-th/0001041}}].

\bibitem{Aharony:1997an}
O.~Aharony, M.~Berkooz, and N.~Seiberg, {\it {Light cone description of (2,0)
  superconformal theories in six-dimensions}},  {\em Adv.Theor.Math.Phys.} {\bf
  2} (1998) 119--153, [\href{http://xxx.lanl.gov/abs/hep-th/9712117}{{\tt
  hep-th/9712117}}].

\bibitem{Bouwknegt:1992wg}
P.~Bouwknegt and K.~Schoutens, {\it {W symmetry in conformal field theory}},
  {\em Phys.Rept.} {\bf 223} (1993) 183--276,
  [\href{http://xxx.lanl.gov/abs/hep-th/9210010}{{\tt hep-th/9210010}}].

\bibitem{Tseytlin:2000sf}
A.~A. Tseytlin, {\it {R**4 terms in 11 dimensions and conformal anomaly of
  (2,0) theory}},  {\em Nucl.Phys.} {\bf B584} (2000) 233--250,
  [\href{http://xxx.lanl.gov/abs/hep-th/0005072}{{\tt hep-th/0005072}}].

\bibitem{Harvey:1998bx}
J.~A. Harvey, R.~Minasian, and G.~W. Moore, {\it {NonAbelian tensor multiplet
  anomalies}},  {\em JHEP} {\bf 9809} (1998) 004,
  [\href{http://xxx.lanl.gov/abs/hep-th/9808060}{{\tt hep-th/9808060}}].

\bibitem{Intriligator:2000eq}
K.~A. Intriligator, {\it {Anomaly matching and a Hopf-Wess-Zumino term in 6d,
  N=(2,0) field theories}},  {\em Nucl.Phys.} {\bf B581} (2000) 257--273,
  [\href{http://xxx.lanl.gov/abs/hep-th/0001205}{{\tt hep-th/0001205}}].

\bibitem{Gadde:2011uv}
A.~Gadde, L.~Rastelli, S.~S. Razamat, and W.~Yan, {\it {Gauge Theories and
  Macdonald Polynomials}},  {\em Commun.Math.Phys.} {\bf 319} (2013) 147--193,
  [\href{http://xxx.lanl.gov/abs/1110.3740}{{\tt arXiv:1110.3740}}].

\bibitem{Pope:1989ew}
C.~Pope, L.~Romans, and X.~Shen, {\it {The Complete Structure of W(Infinity)}},
   {\em Phys.Lett.} {\bf B236} (1990) 173.

\bibitem{Pope:1989sr}
C.~Pope, L.~Romans, and X.~Shen, {\it {$W$(infinity) and the Racah-wigner
  Algebra}},  {\em Nucl.Phys.} {\bf B339} (1990) 191--221.

\bibitem{FigueroaO'Farrill:1992cv}
J.~M. Figueroa-O'Farrill, J.~Mas, and E.~Ramos, {\it {A One parameter family of
  Hamiltonian structures for the KP hierarchy and a continuous deformation of
  the nonlinear W(KP) algebra}},  {\em Commun.Math.Phys.} {\bf 158} (1993)
  17--44, [\href{http://xxx.lanl.gov/abs/hep-th/9207092}{{\tt
  hep-th/9207092}}].

\bibitem{Khesin:1993ru}
B.~Khesin and I.~Zakharevich, {\it {Poisson - Lie group of pseudodifferential
  symbols}},  {\em Commun.Math.Phys.} {\bf 171} (1995) 475--530,
  [\href{http://xxx.lanl.gov/abs/hep-th/9312088}{{\tt hep-th/9312088}}].

\bibitem{Khesin:1993ww}
B.~Khesin and I.~Zakharevich, {\it {Poisson Lie group of pseudodifferential
  symbols and fractional KP - KdV hierarchies}},
  \href{http://xxx.lanl.gov/abs/hep-th/9311125}{{\tt hep-th/9311125}}.

\bibitem{Gaberdiel:2011wb}
M.~R. Gaberdiel and T.~Hartman, {\it {Symmetries of Holographic Minimal
  Models}},  {\em JHEP} {\bf 1105} (2011) 031,
  [\href{http://xxx.lanl.gov/abs/1101.2910}{{\tt arXiv:1101.2910}}].

\bibitem{Campoleoni:2011hg}
A.~Campoleoni, S.~Fredenhagen, and S.~Pfenninger, {\it {Asymptotic W-symmetries
  in three-dimensional higher-spin gauge theories}},  {\em JHEP} {\bf 1109}
  (2011) 113, [\href{http://xxx.lanl.gov/abs/1107.0290}{{\tt
  arXiv:1107.0290}}].

\bibitem{D'Hoker:1999ea}
E.~D'Hoker, D.~Z. Freedman, S.~D. Mathur, A.~Matusis, and L.~Rastelli, {\it
  {Extremal correlators in the AdS / CFT correspondence}},
  \href{http://xxx.lanl.gov/abs/hep-th/9908160}{{\tt hep-th/9908160}}.

\bibitem{Gadde:2009kb}
A.~Gadde, E.~Pomoni, L.~Rastelli, and S.~S. Razamat, {\it {S-duality and 2d
  Topological QFT}},  {\em JHEP} {\bf 1003} (2010) 032,
  [\href{http://xxx.lanl.gov/abs/0910.2225}{{\tt arXiv:0910.2225}}].

\bibitem{Gaiotto:2012xa}
D.~Gaiotto, L.~Rastelli, and S.~S. Razamat, {\it {Bootstrapping the
  superconformal index with surface defects}},
  \href{http://xxx.lanl.gov/abs/1207.3577}{{\tt arXiv:1207.3577}}.

\bibitem{Lemos:2012ph}
M.~Lemos, W.~Peelaers, and L.~Rastelli, {\it {The Superconformal Index of Class
  S Theories of Type D}},  \href{http://xxx.lanl.gov/abs/1212.1271}{{\tt
  arXiv:1212.1271}}.

\bibitem{Mekareeya:2012tn}
N.~Mekareeya, J.~Song, and Y.~Tachikawa, {\it {2d TQFT structure of the
  superconformal indices with outer-automorphism twists}},  {\em JHEP} {\bf
  1303} (2013) 171, [\href{http://xxx.lanl.gov/abs/1212.0545}{{\tt
  arXiv:1212.0545}}].

\bibitem{forthcoming_classS}
C.~Beem, W.~Peelaers, L.~Rastelli, and B.~C. van Rees, {\it {Chiral algebras of
  class S}},  \href{http://xxx.lanl.gov/abs/1408.6522}{{\tt arXiv:1408.6522}}.

\bibitem{Arakawa:2007aa}
T.~Arakawa, {\it {Characters of representations of affine Kac-Moody Lie
  algebras at the critical level}},
  \href{http://xxx.lanl.gov/abs/0706.1817}{{\tt arXiv:0706.1817}}.

\bibitem{Gunaydin:1985tc}
M.~Gunaydin and N.~Warner, {\it {Unitary Supermultiplets of Osp(8/4,r) and the
  Spectrum of the S(7) Compactification of Eleven-dimensional Supergravity}},
  {\em Nucl.Phys.} {\bf B272} (1986) 99.

\bibitem{Dolan:2005wy}
F.~Dolan, {\it {Character formulae and partition functions in higher
  dimensional conformal field theory}},  {\em J.Math.Phys.} {\bf 47} (2006)
  062303, [\href{http://xxx.lanl.gov/abs/hep-th/0508031}{{\tt
  hep-th/0508031}}].

\bibitem{Bianchi:2006ti}
M.~Bianchi, F.~Dolan, P.~Heslop, and H.~Osborn, {\it {N=4 superconformal
  characters and partition functions}},  {\em Nucl.Phys.} {\bf B767} (2007)
  163--226, [\href{http://xxx.lanl.gov/abs/hep-th/0609179}{{\tt
  hep-th/0609179}}].

\end{thebibliography}\endgroup

\end{document}